\documentclass[a4paper, 11pt]{article}
\usepackage[]{graphicx}
\usepackage[dvipsnames,table]{xcolor}
\usepackage{alltt}
\PassOptionsToPackage{hyphens}{url}\usepackage{hyperref}
\usepackage{amsmath, amssymb}
\usepackage{doi} 
\usepackage[round, authoryear]{natbib} 
\usepackage{multirow}
\usepackage{times}
\usepackage{comment}
\usepackage{caption}
\usepackage{subcaption}
\usepackage{enumitem} 
\usepackage{tabularx}
\usepackage{bbm}
\usepackage{algorithm}
\usepackage{algpseudocode}
\usepackage{microtype}
\usepackage{booktabs} 
\usepackage[title]{appendix} 
\usepackage{nameref} 
\usepackage[dvipsnames,table]{xcolor}
\usepackage[onehalfspacing]{setspace} 
\usepackage[labelfont=bf,font=small]{caption} 
\usepackage{newtxtext,newtxmath}
\usepackage{sectsty} 
\allsectionsfont{\sffamily} 
\usepackage[labelfont={bf,sf},font=small]{caption} 
\usepackage{orcidlink} 
\usepackage{fancyhdr}
\usepackage{geometry}
\makeatletter
\def\maxwidth{ %
  \ifdim\Gin@nat@width>\linewidth
    \linewidth
  \else
    \Gin@nat@width
  \fi
}
\makeatother

\usepackage{framed}
\makeatletter
 {\par\unskip\endMakeFramed%
 \at@end@of@kframe}
\makeatother

\definecolor{unipd_red}{HTML}{b40908}
\newcommand\mail{roberto.macridemartino@deams.units.it}
\title{\vspace{-2em}
  \textbf{\textbf{Eliciting prior information from clinical trials via calibrated Bayes factor}}
}
\author{
   \textbf{Roberto Macrì Demartino}\thanks{Corresponding author e-mail: \href{mailto:\mail}{\texttt{\mail}}} \textsuperscript{ $a$}   \orcidlink{0000-0002-5296-6566}, \textbf{Leonardo Egidi}\textsuperscript{$a$} \orcidlink{0000-0003-3211-905X}, \textbf{Nicola Torelli}\textsuperscript{$a$}  \orcidlink{0000-0001-9523-5336}, and \textbf{Ioannis Ntzoufras}\textsuperscript{$b$}  \orcidlink{0000-0002-7615-0334}  \\
  \small\textsuperscript{$a$} Department of Economics, Business, Mathematics and Statistics ``Bruno de Finetti", University of Trieste,\\
  \small Via A. Valerio 4/1, Trieste, 34127, Italy \\
  \small\textsuperscript{$b$} Department of Statistics, Athens University of Economics and Business, Trias 2, Athens, 11362, Greece 
}
\date{}

\IfFileExists{upquote.sty}{\usepackage{upquote}}{}

\usepackage{hyperref}
\hypersetup{
  bookmarksopen=true,
  breaklinks=true,
  colorlinks=true,
  linkcolor=blue,
  anchorcolor=black,
  citecolor=blue,
  urlcolor=black,
}

\begin{document}

\maketitle

\begin{abstract}
In the Bayesian framework power prior distributions are increasingly adopted in clinical trials and similar studies to incorporate external and past information, typically to inform the parameter associated with a treatment effect. Their use is particularly effective in scenarios with small sample sizes and where robust prior information is actually available. A crucial component of this methodology is represented by its weight parameter, which controls the volume of historical information incorporated into the current analysis. This parameter can be considered as either fixed or random. Although various strategies exist for its determination, eliciting the prior distribution of the weight parameter according to a full Bayesian approach remains a challenge. In general, this parameter should be carefully selected to accurately reflect the available prior information without dominating the posterior inferential conclusions. To this aim, we propose a novel method for eliciting the prior distribution of the weight parameter through a simulation-based calibrated Bayes factor procedure. This approach allows for the prior distribution to be updated based on the strength of evidence provided by the data: The goal is to facilitate the integration of historical data when they align with current information and to limit it when discrepancies arise in terms, for instance, of prior-data conflicts. The performance of the proposed method is tested through simulation studies and applied to real data from clinical trials.
  \\[1ex]
\end{abstract}

\section{Introduction}
\label{sec:1}
In recent years, biostatistical applications are commonly characterized by insufficient sample sizes, which are crucial for accurate parameter estimation. Meanwhile, in the clinical framework, a large amount of historical or related data are often available. This has led to a growing interest in the use of historical data to enhance the design and analysis of new studies, particularly in clinical trials where recruiting patients can be ethically challenging. Notably, the sequential nature of information updating has made Bayesian approaches with informative priors particularly popular in this context \citep[among others]{chen_ibrahim2000, spiegelhalter2004bayesian, neuenschwander2010summarizing, hobbs2011hierarchical, Hobbs2012, viele2014use, Schmidli_2014, Yang_2023, Alt_2024}. These methods facilitate the incorporation of historical data into the analysis by eliciting informative priors on the model parameters, thereby improving the robustness and efficiency of statistical inference. However, the elicitation of informative priors is widely recognized as a complex process because of the challenges in quantifying and synthesizing prior information into suitable prior distributions. For an in-depth and comprehensive analysis of the topic, see \citet{spiegelhalter2004bayesian} and \citet{Neuenschwander2020}. Consequently, there is a pressing need to develop more efficient methods for synthesizing and quantifying prior information \citep{Ibrahim2015}. Specifically, there is a growing concern about the adaptive incorporation of historical data, especially in the presence of data heterogeneity and rapid changes in initial trial conditions \citep{ollier2020}.

In this framework, the power prior \citep{chen_ibrahim2000} is a popular method that allows historical data to influence the prior distribution in a flexible and controlled way. \citet{Ibrahim2003} provided a formal justification for power priors, demonstrating their effectiveness as a valuable class of informative priors. This effectiveness stems from their ability to minimize a convex sum of Kullback-Leibler (KL) divergences between two distinct posterior densities, in which one does not include any historical data, whereas the other fully integrates this information into the current analysis. Additionally, \citet{DeSantis_2006} proposed further operational justifications, linking them to the so-called geometric priors. Notably, power priors have been employed across a broad spectrum of models including generalized linear models (GLMs), generalized linear mixed models (GLMMs), and survival models \citep{ChenIbrahimShao2000, ChenIbrahim2006, Ibrahim2015}. At its core, the idea is to raise the likelihood of the historical data to a weight parameter $\delta$, usually defined between zero and one. This scalar parameter plays a crucial role in the power prior methodology as it determines the degree to which historical data influence the prior distribution. Specifically, when $\delta$ is set to zero, the power prior completely discounts historical information; conversely, setting $\delta$ to one fully integrate historical information into the prior.

As is intuitive, the role that the weight parameter $\delta$ plays in the final inferential conclusions is not negligible. Thus, several strategies have been developed to specify the weight parameter $\delta$, treating it either as a fixed or a random quantity. If fixed in advance, the weight parameter $\delta$ can be set based on prior knowledge or through sensitivity analysis, considering specific criteria for borrowing information based on the prior-data conflict \citep{evans2006, egidi_2022}. If treated as random, an initial prior distribution -- typically a Beta distribution -- is assigned to $\delta$, and the use of the joint normalized power prior \citep{duan2006, Neuenschwander2009} is recommended. Notably, \citet{Nikolakopoulos2018} introduced a method to estimate this parameter using predictive distributions, aiming to control type I error by calibrating to the degree of similarity between current and historical data. Furthermore, \citet{liu2018} recommended setting $\delta$ through a dynamic \textit{p}-value, assessing the compatibility of current and historical data based on the test‐then‐pool methodology. \citet{Gravestock2017, Gravestock_2019} proposed an empirical Bayes-type approach to estimate the weight parameter by maximizing the marginal likelihood. \citet{Bennett2021} suggested two novel approaches for binary data, focusing on equivalence probability and tail area probabilities. \citet{mariani2023} explored the use of the Hellinger distance to compare the posterior distributions of the control parameter from current and historical data, respectively.
These techniques provide valuable insights on determining a specific fixed value for $\delta$. However, in a fully Bayesian context, only \citet{shen2023optimal} has developed methods to specify the shape parameters of a Beta initial prior for $\delta$, using two minimization criteria: Kullback-Leibler (KL) divergence and mean squared error (MSE).
Therefore, in a fully Bayesian framework, eliciting the initial distribution of the weight parameter controlling the amount of historical information remains a challenging and underexplored area.

This paper aims to propose a novel Bayesian algorithmic approach for eliciting the initial prior distribution for $\delta$ in a somehow optimal way. This involves the use of a simulation-based calibrated Bayes factor, employing hypothetical replications generated from the posterior predictive distribution, to compare competing prior distributions for $\delta$. Following the approach in \citet{shen2023optimal}, a well-balanced prior should promote the integration of historical data when there is agreement with the current information and limit this integration when discrepancies arise between the two datasets. 

The paper is organized as follows. Section \ref{Power prior} provides a review of the power prior methodology, discussing the use of the weight parameter as both a fixed and a random quantity. Furthermore, Section \ref{Calibrated BF} illustrates the proposed calibrated Bayes factor to elicit a well-balanced initial prior distribution for $\delta$. Sections \ref{Simulation Study} and \ref{Cancer Study} explore the proposed methodology through simulation studies and real data analysis from two clinical trials, E2696 and E1694, that investigated the effectiveness of interferon in melanoma treatment \citep{kirkwood2001E2696, kirkwood2001E1694}. Finally, Section \ref{Discussion} provides concluding remarks about the discussed method, highlighting its strengths and limitations, and provides insights into future developments.

\section{Power priors} \label{Power prior}

Power priors have become increasingly popular for the development of informative priors, especially in clinical trials where past information is often available. These priors effectively integrate knowledge from historical data into the specification of informative priors. Let consider $\boldsymbol{\theta}$ as the vector parameter of interest in the model, and let $y_0$ represent the historical data with their corresponding likelihood function denoted by $L(\boldsymbol{\theta} \mid y_0)$. The basic formulation of the power prior \citep{chen_ibrahim2000} is given by
\begin{equation}
    \pi(\boldsymbol{\theta} \mid y_0, \delta) \propto L(\boldsymbol{\theta} \mid y_0)^\delta \pi_0(\boldsymbol{\theta}),
    \label{eqn: 1}
\end{equation}
where $\delta \in [0,1]$ is the scalar weight parameter, and $\pi_0(\boldsymbol{\theta})$ is the initial -- often non-informative -- prior for $\boldsymbol{\theta}$. This model can be seen as a generalization of the classical Bayesian updating of $\pi_0(\boldsymbol{\theta})$. Additionally, as noted by \citet{Ibrahim2015}, the parameter $\delta$ plays a crucial role in determining the shape of the prior distribution for $\boldsymbol{\theta}$. Furthermore, updating the power prior in (\ref{eqn: 1}) with the current data likelihood $L(\boldsymbol{\theta} \mid y)$ yields the following posterior distribution of $\boldsymbol{\theta}$
\begin{equation*}
    \pi(\boldsymbol{\theta} \mid y, y_0, \delta) \propto L(\boldsymbol{\theta} \mid y) L(\boldsymbol{\theta} \mid y_0)^\delta \pi_0(\boldsymbol{\theta}).
\end{equation*}

The formulation in (\ref{eqn: 1}) is conditional on the weight parameter and requires a predetermined and known value for $\delta$. Therefore, to ensure an appropriate level of historical information borrowing while managing prior-data conflict, sensitivity analysis is recommended. In addition to the dynamic methods outlined in the introduction, \citet{Ibrahim2015} also proposed several other statistical methods, including the Penalized Likelihood-type Criterion (PLC), Marginal Likelihood Criterion (MLC), Deviance Information Criterion (DIC), and the Logarithm of the Pseudo-Marginal Likelihood (LPML) Criterion.

\subsection{Hierarchical power priors}
A natural extension of the power prior in (\ref{eqn: 1}) can be achieved by accounting for uncertainty about the weight parameter $\delta$. This involves adopting a hierarchical power prior where $\delta$ is treated as a random variable. To achieve this, a Beta prior distribution is assigned to $\delta$, leading to the so-called joint unnormalized power prior \citep{chen_ibrahim2000} for both $\boldsymbol{\theta}$ and $\delta$
\begin{equation}
\begin{aligned}
    \pi(\boldsymbol{\theta}, \delta \mid y_0)  &= \pi(\boldsymbol{\theta} \mid y_0, \delta)\pi_0(\delta) 
    \\
    &\propto L(\boldsymbol{\theta} \mid y_0)^\delta \pi_0(\boldsymbol{\theta}) \pi_0(\delta),    
\end{aligned}
    \label{eqn: 2}
\end{equation}
where $\pi_0(\boldsymbol{\theta})$ and $\pi_0(\delta)$ represent the initial priors for $\boldsymbol{\theta}$ and $\delta$, respectively. 

However, as noted by \citet{duan2006} and \citet{Neuenschwander2009}, the formulation in (\ref{eqn: 2}) lacks a normalizing constant, leading to potential inconsistencies in the joint posterior distributions for $(\boldsymbol{\theta}, \delta)$ derived from different forms of likelihood functions, such as those based on raw data versus those based on the distribution of sufficient statistics \citep{YE_2022}. Consequently, \citet{duan2006} proposed a normalized power prior (NPP) which involves first setting a conditional prior on $\boldsymbol{\theta}$ given $\delta$, followed by an initial prior distribution for $\delta$. 
The resulting joint normalized power prior for $(\boldsymbol{\theta}, \delta)$ is  

\begin{equation}
\begin{aligned}
    \pi\left(\boldsymbol{\theta}, \delta \mid y_0\right) &= \pi(\boldsymbol{\theta}\mid y_0,\delta)\pi_0(\delta) \\
    &= \frac{L\left(\boldsymbol{\theta} \mid y_0\right)^\delta \pi_0(\boldsymbol{\theta}) }{C(\delta)} \pi_0(\delta),
    \label{eqn: pp_normalized}
\end{aligned}
\end{equation}
where the normalizing constant $C(\delta)$ is
\begin{equation}
    C(\delta) = \int_{\Theta} L(\boldsymbol{\theta}\mid y_0)^\delta \pi_0(\boldsymbol{\theta})\mathrm{d}\boldsymbol{\theta}.
     \label{eqn: Norm_Const}
\end{equation}
In light of the current data, the joint posterior distribution is
\begin{equation}
\begin{aligned}
    \pi\left(\boldsymbol{\theta}, \delta \mid y, y_0\right) &= \frac{L(\boldsymbol{\theta}\mid y) \pi\left(\boldsymbol{\theta}, \delta \mid y_0\right)}{\int_0^1 \int_\Theta L(\boldsymbol{\theta}\mid y) \pi\left(\boldsymbol{\theta}, \delta \mid y_0\right) \mathrm{d}\boldsymbol{\theta}\mathrm{d}\delta} \\
    &\propto  L(\boldsymbol{\theta}\mid y) \frac{L\left(\boldsymbol{\theta} \mid y_0\right)^\delta \pi_0(\boldsymbol{\theta}) }{C(\delta)}\pi_0(\delta).
    \label{eqn: pp_normalized_post}
\end{aligned}
\end{equation}
The marginal posterior distribution of $\delta$ is
\begin{equation*}
\begin{aligned}
    \pi\left(\delta \mid y, y_0\right)  \propto   \frac{\pi_0(\delta)}{C(\delta)}\int_{\Theta}L(\boldsymbol{\theta}\mid y) L\left(\boldsymbol{\theta} \mid y_0\right)^\delta \pi_0(\boldsymbol{\theta}) \mathrm{d}\boldsymbol{\theta}.
    \label{eqn: pp_normalized_post_delta}
\end{aligned}
\end{equation*}
Integrating $\delta$ out in (\ref{eqn: pp_normalized_post}), the marginal posterior of $\boldsymbol{\theta}$ can be written as
\begin{equation*}
\begin{aligned}
    \pi\left(\boldsymbol{\theta} \mid y, y_0\right)  \propto   \pi_0(\boldsymbol{\theta})L(\boldsymbol{\theta}\mid y)\int_0^1\frac{L(\boldsymbol{\theta}\mid y_0)^{\delta} \pi_0(\delta)}{C(\delta)}\mathrm{d}\delta.
    \label{eqn: pp_normalized_post_theta}
\end{aligned}
\end{equation*}

The joint power prior framework offers the advantage of incorporating uncertainty regarding the weight parameter $\delta$ into the power prior formulation. This approach allows the data to determine the appropriate weight for historical information based on its compatibility with current observations. Furthermore, explicitly accounting for this uncertainty increases the flexibility in modeling the agreement between historical and current data.

In addition, a crucial theoretical advantage of the joint normalized power prior with respect to the formulation in (\ref{eqn: 2}) is its adherence to the likelihood principle. This ensures that the posterior distributions in (\ref{eqn: pp_normalized_post}) accurately reflect the compatibility between current and historical data. Furthermore, this approach has further theoretical justification, as the power prior formulation in (\ref{eqn: pp_normalized}) is shown to minimize the weighted KL divergence \citep{YE_2022}.

From a computational perspective, the additional effort required for the normalized power prior compared to the unnormalized power prior involves computing the normalizing constant in (\ref{eqn: Norm_Const}). For certain models, this integral can be solved analytically, resulting in a closed-form expression for the joint posterior as specified in (\ref{eqn: pp_normalized_post}). However, for more complex models, the normalizing constant $C(\delta)$ must be determined numerically. Consequently, the posterior distribution in (\ref{eqn: pp_normalized_post}) falls into the category of the doubly intractable distributions \citep{Carvalho_2021}, and numerical methods such as Markov Chain Monte Carlo (MCMC) \citep{Robert2004} are required.

\section{The calibrated Bayes factor} \label{Calibrated BF}

Eliciting a well-balanced initial prior distribution for $\delta$ has proven to be challenging. Intuitively, this prior should encourage borrowing when the data are compatible and limit borrowing when they are in conflict. In this section, we propose a calibrated Bayes factor, hereafter CBF, that is a simulation-based algorithmic technique designed to effectively discriminate between some competing initial Beta prior distributions for $\delta$.
The proposed CBF aims to provide more robust decisions about which initial Beta prior for $\delta$ may be used. 
Specifically, this involves analyzing the behaviour of the Bayes factor  \citep{jeffreys1961theory, kass1993,kass1995}, henceforth BF, using different hypothetical replications generated from the posterior predictive distributions, while assessing how surprising the observed value of the BF is.  
\subsection{The Bayes factor}
The BF provides a general Bayesian method to assess the relative evidence in support of competing hypotheses based on their compatibility with the observed data. Furthermore, the BF represents the ratio between the posterior and the prior odds when comparing two distinct point hypotheses. Specifically, let consider two competing hypotheses about the initial prior distribution of the weight parameter
$$
 \mathcal{H}_0: \delta \sim \mathrm{Beta}(\eta_0,\nu_0)\; \text{  vs.  } \; \mathcal{H}_1 :\delta \sim \mathrm{Beta}(\eta_1,\nu_1),  
$$
  where $\eta_i$ and $\nu_i$ represent the strictly positive Beta shape parameters under the hypothesis $\mathcal{H}_i$, for $i = \left\{0,1\right\}$. The corresponding BF can be expressed as the ratio of the two marginal likelihoods
\begin{equation}
\begin{aligned}
    \mathrm{BF}_{0,1}(y) &= \dfrac{m(y \mid \mathcal{H}_1)}{m(y \mid \mathcal{H}_0)},
\end{aligned}
    \label{eqn:bf}
\end{equation}
where the marginal likelihood is
\begin{align*}
m(y \mid \mathcal{H}_i) =  \int_0^1 \dfrac{\int_{\Theta} L\left(\boldsymbol{\theta} \mid y\right) L\left(\boldsymbol{\theta} \mid y_0\right)^\delta \pi_0(\boldsymbol{\theta}) \pi_{0}(\delta \mid \mathcal{H}_i) \mathrm{d} \boldsymbol{\theta}}{\int_{\Theta} L\left(\boldsymbol{\theta} \mid y_0\right)^\delta \pi_0(\boldsymbol{\theta}) \mathrm{d} \boldsymbol{\theta}}   \mathrm{d} \delta,    
\end{align*}
with $\pi_{0}(\delta \mid \mathcal{H}_{i})$ representing the initial Beta prior distribution for $\delta$ under the hypothesis $\mathcal{H}_i$, for $i = \left\{0,1\right\}$. 
Assuming equal prior probabilities for both models, a BF exceeding one suggests stronger evidence in favor of $\mathcal{H}_1$. Conversely, a BF below one denotes stronger evidence for $\mathcal{H}_0$. A BF close to one indicates no clear preference for either hypothesis, reflecting a similar degree of empirical evidence for both $\mathcal{H}_0$ and $\mathcal{H}_1$. Several approaches have been developed to summarize and classify the strength of evidence according to the observed BF. Firstly, \citet{jeffreys1961theory} introduced a categorization, as illustrated in Table \ref{table:Jeffreys scale}. Subsequently, \citet{kass1995} streamlined this scale by omitting one category and redefining the thresholds. 
Lastly, \citet[Table~7.1, p.105]{lee2014bayesian} further refined Jeffreys' scale with additional modifications.

For most complex models, the BF computation is challenging since the marginal likelihood is not analytically tractable. Therefore, numerical approximation methods become essential. A widely used algorithm is the so-called bridge sampling \citep{meng1996simulating}. This method employs a Monte Carlo technique, generating samples from an auxiliary distribution that bridges the model's prior and posterior distribution. The generated samples are then used to calculate bridge sampling weights, which correct the bias introduced by the auxiliary distribution, providing an unbiased estimate of the marginal likelihood.

Another notable method is the Savage–Dickey algorithm \citep{Dickey1970}. This method approximates the BF by calculating the ratio of the posterior and prior densities at a model parameter value of zero.  However, its use is limited to nested models and may be unstable if the posterior density significantly deviates from zero.

\begin{table}[h!]
\begin{center}
\caption{Scale of evidence for the BF proposed by \citet{jeffreys1961theory}.}
\label{table:Jeffreys scale}
\begin{tabular}{ccc}
\toprule
\(\mathrm{BF}_{0,1}\) & \(\log_{10}(\mathrm{BF}_{0,1})\) & Evidence Category \\
\midrule
\(1\) - \(3.16\) & \(0\) - \(0.5\) & Barely worth mentioning \\
\(3.16\) - \(10\) & \(0.5\) - \(1\) & Substantial evidence for $\mathcal{H}_1$ \\
\(10\) - \(31.62\) & \(1\) - \(1.5\) & Strong evidence for $\mathcal{H}_1$\\
\(31.62\) - \(100\) & \(1.5\) - \(2\) & Very strong evidence for $\mathcal{H}_1$\\
\(> 100\) & \(> 2\) & Decisive evidence for $\mathcal{H}_1$ \\
\bottomrule
\end{tabular}
\end{center}
\end{table}

\subsection{Simulation-based calibration }

The BF has an inherent dependence on the observed data when used for decision-making. Consequently, decisions based solely on current observations may lead to potentially misleading conclusions due to the possible fluctuations and noise present in the data. Furthermore, \citet{schad2023workflow} emphasized two crucial issues regarding BF computations: the instability of BF estimates in complex statistical models and the potential bias within these estimates. Therefore, to effectively and responsibly employ the BF, it is crucial to adjust and calibrate it, ensuring that the conclusions drawn are more robust and reliable.

The concept of simulation-based calibration, hereafter SBC, was originally developed to validate the computational correctness of applied Bayesian inference methods \citep[among others]{geweke2004, cook2006, talts2018validating, gelman2020bayesian, Schad2021, modrak2023}. In addition, \citet{schad2023workflow} proposed a structured approach based on SBC to verify the accuracy of the BF calculations. Their calibration method involves simulating multiple artificial datasets to assess whether a BF estimated in a given analysis is accurate or biased. This type of calibration approach is intuitive and logical, as it mirrors the classical approach of hypothesis testing, where the decision-making criterion is determined by the sampling distribution under the hypothesis of repeated sampling.

The SBC-inspired method explored in this paper \color{black} is motivated by the insights of \citet{Garcia2005}, who posited that BFs should be considered as a random variable before observing the sample. This perspective emphasizes the necessity of calibrating the BF to accurately account for the inherent randomness in the data. However, the analytic form of the BF distribution is frequently not available. In such cases, a simulation-intensive approach becomes a valuable tool to approximate this distribution \citep{VLACHOS2003223}. This involves generating replicated datasets from some type of predictive distribution or other data-generating processes, estimating the marginal likelihoods with these synthetic datasets, and then computing the BFs. By iterating this process multiple times, an approximation of the BF distribution is obtained. Subsequently, once the data have been collected, the observed BF can be used as a measure of agreement between the observed data and the underlying statistical model. 

Both \citet{Garcia2005} and \citet{schad2023workflow} suggested a calibration method based on the prior predictive distribution.  However, this approach is not suitable in our context due to the inherent bias in the replicated data from the prior predictive distribution toward the historical data $y_0$, as highlighted by the structure of Equation (\ref{eqn: pp_normalized}). This can potentially yield replicated samples that are much more in agreement with the historical data than the historical data are with the current data. Consequently, we explored the use of hypothetical replications generated from the posterior predictive distribution to approximate the distribution of the BF; where the predictive distribution is given by
    \begin{align*}
        p\left(y^{\text {rep }} \mid y\right)=\int_\Theta L\left(y^{\text {rep }} \mid \boldsymbol{\theta}\right)  \pi(\boldsymbol{\theta} \mid y, y_0, \delta) \mathrm{d} \boldsymbol{\theta}.
    \end{align*}
A key advantage of this approach is the use of the information in the current data $y$ through the posterior distribution, focusing on a relevant region within the parameter space \citep{ROBERT2022103}. 
Furthermore, the BF computed using the replications from the posterior predictive distribution $\mathrm{BF}_{0,1}(y^{\text{rep}})$, henceforth the replicated BF, mimics the behavior of the BF using the original data $y$ when a specific model is the true data-generating mechanism. Our CBF approach aims to define a decision criterion that not only assesses the inherent variability of the BF but also incorporates the observed data. Thus, to effectively ensure a more comprehensive and balanced decision rule, it is essential to define a criterion that incorporates the observed BF, denoted by $\mathrm{BF}_{0,1}(y)$, as a measure of surprise, favoring scenarios where it is less unexpected. 
We propose a criterion based on 
\begin{itemize}
    \item The survival function of the BF distribution, denoted by $S_{\mathrm{BF}_{0,1}}(\cdot)$.
    \item The inclusion of the observed BF within a defined Highest Posterior Density Interval (HPDI).
\end{itemize}
Specifically, the decision criterion focuses on selecting alternative hypotheses that provide stronger evidence against the null hypothesis. This is achieved by giving preference to the BF distribution that yields more values in favor of the alternative hypothesis. This implies that the survival function of the BF distribution, calculated at the value where the BF indicates equal support for both hypotheses, is greater than 0.5.
Furthermore, the inclusion of the observed BF within a defined HPDI assesses its coherence with respect to the underlying distribution.
This dual approach ensures a balanced and comprehensive decision rule, accounting for both the BF distribution and the surprise measure associated with the observed BF.

\subsection{The procedure} \label{subsec: The procedure}

To streamline our method, we focus on the logarithmic transformation of the BF, referred to as Log-BF. This transformation is advantageous because values less than zero suggest evidence for the null hypothesis $\mathcal{H}_0$, while values greater than zero provide evidence for the alternative hypothesis $\mathcal{H}_1$. To further simplify comparisons and improve interpretability, we use a reference null hypothesis $\mathcal{H}_0$ computing the Log-BF between each alternative hypothesis and the reference, reducing the number of comparisons from $\binom{M}{2}$ to $(M-1)$ Log-BFs. The chosen reference null hypothesis is the $\mathrm{Beta}(1,1)$ prior for $\delta$, a commonly used non-informative prior for the weight parameter \citep{Ibrahim2015}. Notably, this choice is motivated by our goal of establishing a more informative prior for the weight parameter compared to the standard uniform prior.

The initial step involves defining a reasonable grid of potential alternative hypotheses related to the Beta initial prior for the weight parameter, denoted by $\boldsymbol{\mathcal{H}} = \left\{\mathcal{H}_1, \ldots, \mathcal{H}_i, \ldots, \mathcal{H}_M\right\}$, with 
\begin{equation}
\label{eqn:alt_hyp}
    \mathcal{H}_i: \delta \sim \mathrm{Beta}(\eta_i, \nu_i),
\end{equation} 
where $\eta_i$ and $\nu_i$ represent the shape parameters under $\mathcal{H}_i$, for  $i=1, \ldots, M$. In particular, the grid should explore parameter space regions ranging from scenarios with minimal borrowing of historical information to those with extensive borrowing. We suggest defining a grid that spans from $0.5$ to $6$, covering a wide range of Beta priors. For instance, a $\mathrm{Beta}(0.5,6)$ prior assigns minimal weight to historical data, while a $\mathrm{Beta}(6,0.5)$ prior incorporates a considerable amount of information from the historical study.

After computing the observed Log-BF between hypothesis $\mathcal{H}_i$ and $\mathcal{H}_0$, the next step involves generating $K$ hypothetical samples from the posterior predictive distribution under each alternative hypothesis $\mathcal{H}_i$, that is $\boldsymbol{y}_{\mathcal{H}_i}^{\mathrm{rep}} = (y_{\mathcal{H}_i, 1}^{\mathrm{rep}}, \ldots, y_{\mathcal{H}_i, K}^{\mathrm{rep}})$, for $i, \ldots, M$. Then, the replicated Log-BF is computed between the alternative hypothesis $\mathcal{H}_i$ and the null hypothesis $\mathcal{H}_0$, for all combinations of $i = 1, \ldots, M$ and $k = 1, \ldots, K$. 

Subsequently, it is essential to define a criterion based on the distribution of the Log-BF, obtained using the replicated Log-BFs, and that incorporates the observed Log-BF as a measure of surprise, favoring scenarios where it is less unexpected. 
Our criterion aims to identify the BF distributions in which the alternative hypothesis $\mathcal{H}_i$, as given in (\ref{eqn:alt_hyp}), is more likely than the null hypothesis $\mathcal{H}_0: \delta \sim \mathrm{Beta}(1,1)$.
This is obtained when the cumulative distribution function (CDF) of the Log-BF at zero is below $0.5$ -- or when the survival function at zero exceeds $0.5$ -- suggesting stronger theoretical evidence in favor of the alternative hypothesis. The observed Log-BF is also incorporated in our decision criterion by considering the hypotheses associated with values greater than zero, reflecting stronger empirical evidence relative to the null hypothesis. The robustness of the observed Log-BF is evaluated by assessing its position within the approximated distribution. Ideally, the observed Log-BF should be within a specific HPDI, indicating that it is not an outlier but rather a value consistent with the underlying Log-BF distribution. Consequently, the well-balanced prior is determined using the following criterion
\begin{equation}
\mathcal{H}_{\mathrm{Opt}} = 
\begin{cases} 
\mathcal{H}^\star & \text{if $\mathcal{H}^\star > 0$} \\
\mathcal{H}_0 &  \text{otherwise},
\end{cases}
\label{eqn:selection_criterion}
\end{equation}
with
\begin{equation*}
  \mathcal{H}^\star =  \operatorname*{arg\,max}_{i=1, \ldots, M}\left\{\mathbbm{1}_{\left\{S_{\log \mathrm{BF}_{0,i}}(0)>0.5 \right\}}S_{\log \mathrm{BF}_{0,i}}(0)\times\mathbbm{1}_{\left\{\log \mathrm{BF}_{0,i}^{\mathrm{obs}}\in (\cdot)\%\mathrm{HPDI}\right\}}\log \mathrm{BF}_{0,i}^{\mathrm{obs}}\right\},
\end{equation*}
where $S_{\log \mathrm{BF}_{0,i}}(0)$ represents the survival function of the Log-BF distribution evaluated at zero, and $\log \mathrm{BF}_{0,i}^{\mathrm{obs}}$ is the observed Log-BF between the alternative hypothesis $\mathcal{H}_i$ and the null hypothesis $\mathcal{H}_0$, for $i = 1, \ldots, M$. The first indicator function focuses on distributions where the alternative hypothesis is more probable than the null hypothesis. That is, selecting distributions where the survival function of the Log-BF at zero is greater than 0.5, indicating higher theoretical evidence for the alternative hypothesis. The second indicator function evaluates the presence of the observed Log-BF within a specific HPDI, working from a Bayesian perspective as a measure of surprise. Furthermore, the computational steps of the CBF procedure are summarized in Algorithm 1.
\begin{algorithm}
\caption{Calibrated Bayes Factor Procedure}
\begin{algorithmic}[1]
\State Define a grid of size $M$ of potential competing hypotheses regarding the Beta prior for the weight parameter $\boldsymbol{\mathcal{H}} = \left\{\mathcal{H}_1, \ldots, \mathcal{H}_i, \ldots, \mathcal{H}_M\right\}$, with $ \mathcal{H}_i: \delta \sim \mathrm{Beta}(\eta_i, \nu_i)$, for $i=1, \ldots, M.$
\State For $i=1,\ldots,M$ compute the Observed Log-BF between hypothesis $\mathcal{H}_i$ and $\mathcal{H}_0$ as in (\ref{eqn:bf}).
\State For $i=1,\ldots,M$ generate $K$ hypothetical samples from the posterior predictive distribution under the alternative hypothesis $\mathcal{H}_i$, that is  $\boldsymbol{y}_{\mathcal{H}_i}^{\mathrm{rep}} = (y_{\mathcal{H}_i, 1}^{\mathrm{rep}}, \ldots, y_{\mathcal{H}_i, K}^{\mathrm{rep}})$.
\State For $i=1,\ldots,M$ and for $k=1,\ldots,K$ compute the replicated Log-BF, $\log\mathrm{BF}_{0,i}(y_{\mathcal{H}_i, k}^{\mathrm{rep}})$, between hypothesis $\mathcal{H}_i$ and $\mathcal{H}_0$ as in (\ref{eqn:bf}) using the hypothetical posterior predictive samples generated in Step 3.
\State Select the Beta prior for $\delta$ using the criterion in (\ref{eqn:selection_criterion}).
\end{algorithmic}
\end{algorithm}

\begin{figure}[htb!]
    \centering
        \includegraphics[width=0.95\textwidth, height=10.5cm]{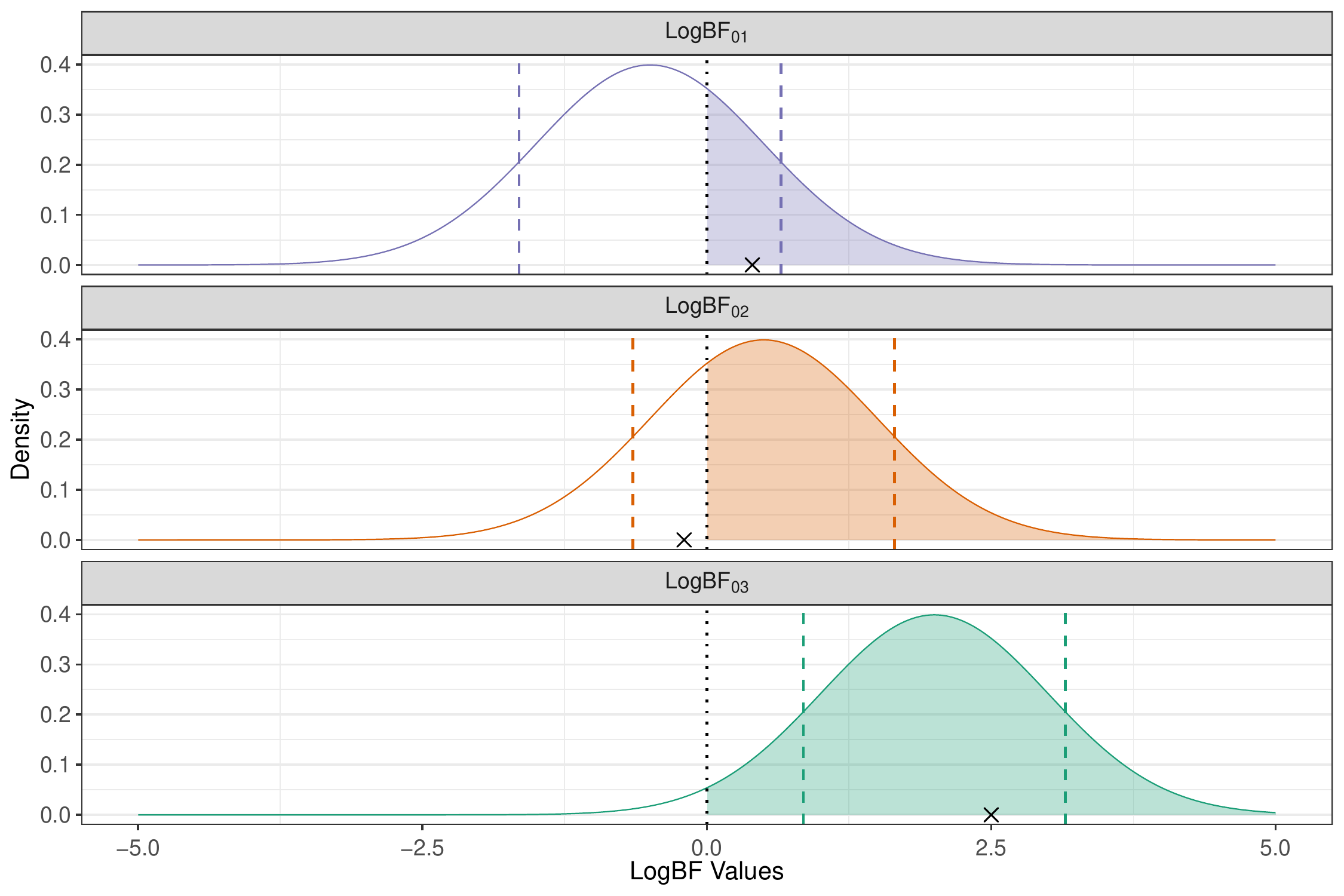}
        \caption{Illustrative example of the CBF procedure. The Log-BF distributions (purple: $\mathcal{H}_1 \;\text{vs.}\; \mathcal{H}_0$, orange: $\mathcal{H}_2 \;\text{vs.}\; \mathcal{H}_0$, green: $\mathcal{H}_3 \;\text{vs.}\; \mathcal{H}_0$), the observed Log-BFs (crossed points) and HPDIs (dashed lines) are shown. The filled area after the zero threshold (dotted line) in each distribution represents the portion where the evidence for the alternative hypothesis is stronger than for the null hypothesis.}
        \label{CBF_procedure}
\end{figure}

Figure \ref{CBF_procedure} provides an illustrative example of the CBF procedure for selecting a well-balanced initial Beta prior for $\delta$. Let consider a null hypothesis $\mathcal{H}_0$ and three alternative hypotheses $\mathcal{H}_i$, for $i = 1,2,3$, regarding the initial Beta prior of $\delta$. The Log-BF distribution for $\mathcal{H}_1$, represented by the purple curve, shows a higher probability of negative Log-BF values, suggesting stronger theoretical evidence in favor of the null hypothesis.  Conversely, the approximated Log-BF distribution for $\mathcal{H}_2$ and $\mathcal{H}_3$, depicted by the orange and green curves, respectively, provides stronger evidence in favor of the associated alternative hypotheses. However, although $\mathcal{H}_2$ shows an observed Log-BF within the selected HPDI, denoted by the dashed lines, the corresponding observed Log-BF value is negative, suggesting empirical evidence in favor of $\mathcal{H}_0$. Only $\mathcal{H}_3$ demonstrates a positive observed Log-BF, which suggests stronger empirical evidence for the alternative hypothesis, but also falls within the respective HPDI. Accordingly, based on the selection criterion in (\ref{eqn:selection_criterion}), a well balanced prior for $\delta$ is the one associated with the alternative hypothesis $\mathcal{H}_3$.

\section{Simulation studies} \label{Simulation Study}
In this section, we assess the efficacy and applicability of the CBF approach through simulation studies. Specifically, we consider three distinct scenarios, each involving different statistical distributions and the corresponding parameters. The main aim of each simulated study is to evaluate the method's ability to identify a well-balanced prior that effectively integrates extensive historical information when current data closely align with the historical data, while minimizing such integration in presence of disagreement. To assess this, we analyze a historical dataset and a series of current datasets that progressively diverge from it. We compare the performance of the proposed CBF approach against other dynamic information borrowing methods, including the NPP with a uniform prior on $\delta$, the robust meta-analytic predictive (RMAP) prior \citep{Schmidli_2014}, and the self-adapting mixture (SAM) prior \citep{Yang_2023}. Notably, the RMAP prior employed assigns equal weight $\omega$ to both the non-informative mixture component and the informative mixture component based on the historical data. As described in Section \ref{subsec: The procedure}, an expanded grid for the Beta prior parameters is used, ranging from $0.5$ to $6$ in increments of $0.5$. 
According to some sensitivity checks, each simulation study considers a $75\%$ HPDI threshold for the selection criterion in (\ref{eqn:selection_criterion}).

\subsection{Poisson log-linear model}
 Let $y_0 = (y_{0,1}, \ldots, y_{0,N_0})$ and $y = (y_1, \ldots, y_N)$ be the count outcome of an historical and a current study, respectively. Let denote with $\boldsymbol{x}_{0,k}=(x_{0,k1}, \ldots, x_{0,kp})$, for $k = 1, \ldots, N_0$, and $\boldsymbol{x}_j=(x_{j1}, \ldots, x_{jp})$, for $j = 1, \ldots, N$, the corresponding covariate vector. 
The Poisson log-linear model is
\begin{align*}
y_j \mid \boldsymbol{\beta} \sim \mathrm{Poisson}(\lambda_j), \quad \text{where } \lambda_j = \exp(\mathbf{x}_j^\top \boldsymbol{\beta}),   
\end{align*}
with $\boldsymbol{\beta}$ being the $p$-dimensional vector of the regression coefficients. Let $\pi_0(\boldsymbol{\beta})$ be the initial multivariate normal prior on $\boldsymbol{\beta}$, the BF is given by
\begin{align*}
    \mathrm{BF}_{0,i}(y) = \frac{\int_0^1 \int L(\boldsymbol{\beta} \mid y, X)[L(\boldsymbol{\beta} \mid y_0, X_0)]^\delta \pi_0(\boldsymbol{\beta})\mathrm{Beta}(\delta \mid \eta_i,\nu_i) \, \mathrm{d}\boldsymbol{\beta} \, \mathrm{d}\delta}{\int_0^1 \int L(\boldsymbol{\beta} \mid y, X)[L(\boldsymbol{\beta} \mid y_0, X_0)]^\delta \pi_0(\boldsymbol{\beta})\mathrm{Beta}(\delta \mid 1,1) \, \mathrm{d}\boldsymbol{\beta} \, \mathrm{d}\delta}.
\end{align*}
where $[L(\boldsymbol{\beta} \mid y_0, X_0)]^\delta = \exp\left(\sum_{k=1}^{N_0} \left[\delta(y_{0,k}\mathbf{x}_{0,k}^\top \boldsymbol{\beta} - \exp(\mathbf{x}_{0,k}^\top \boldsymbol{\beta}))\right]\right)
\prod_{k=1}^{N_0} (y_{0,k}!)^{-\delta}.$
For further details, see Appendix \ref{appendix_poisson}.

\begin{figure}[htb!]
    \centering
        \includegraphics[width=0.95\textwidth, height=8.5cm]{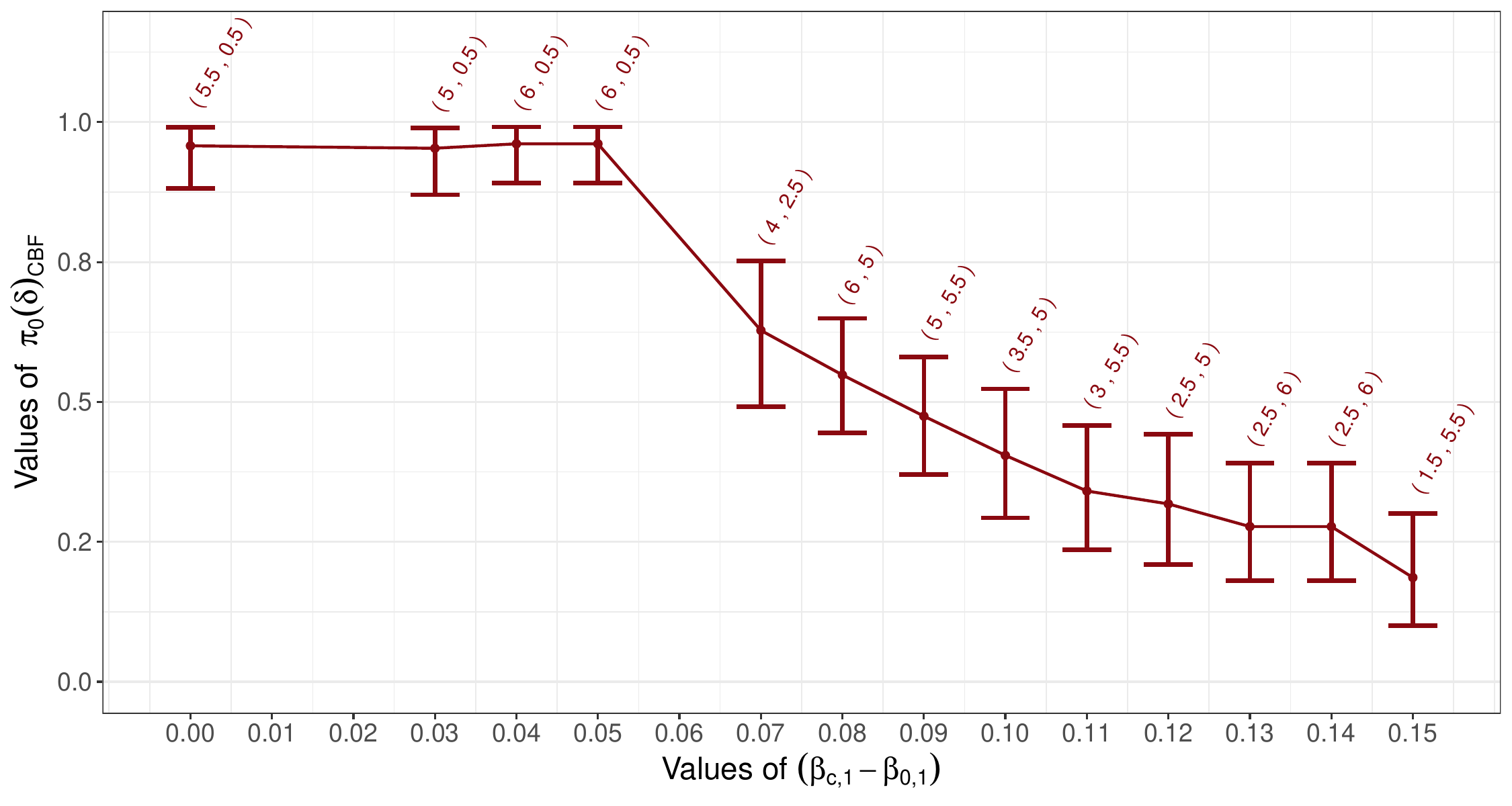}
        \caption{Poisson log-linear model. Shifts in the prior median (point) for the CBF derived prior for $\delta$, considering the 25th and 75th Percentiles (bar), in function of the difference between the current regression coefficient $\beta_{c,1}$ and the historical regression coefficient $\beta_{0,1}$. Above each interval, the corresponding selected Beta parameter are displayed.}
        \label{BF_plot_poisson_opt}
\end{figure}

In this simulation study, we first generate a historical dataset that includes three distinct covariates and an intercept. We then generate a series of current datasets that gradually diverge from the historical data by increasing the regression coefficient $\beta_{c,1}$ associated with the first covariate.
Figure \ref{BF_plot_poisson_opt} illustrates the evolution of the prior selected for $\delta$ in relation to the level of disagreement between current and historical studies. This disagreement is quantified by the difference between the historical and current regression coefficients. 
The plot shows the median values of the selected prior for $\delta$ as points, with error bars representing the first and third quartiles. The values in brackets correspond to the chosen Beta parameters selected from the predefined grid. The plot highlights the procedure's ability to select an appropriate prior according to the level of disagreement between the datasets. As the disagreement grows, the chosen prior shifts from a left-skewed Beta distribution -- a $\mathrm{Beta}(5.5,0.5)$, indicating substantial incorporation of historical data (equal weight to the actual data), to a right-skewed Beta distribution -- a $\mathrm{Beta}(1.5,5.5)$, reflecting more conservative integration of historical information. 

\begin{figure}[htb!]
    \centering
        \includegraphics[width=0.95\textwidth, height=10.5cm]{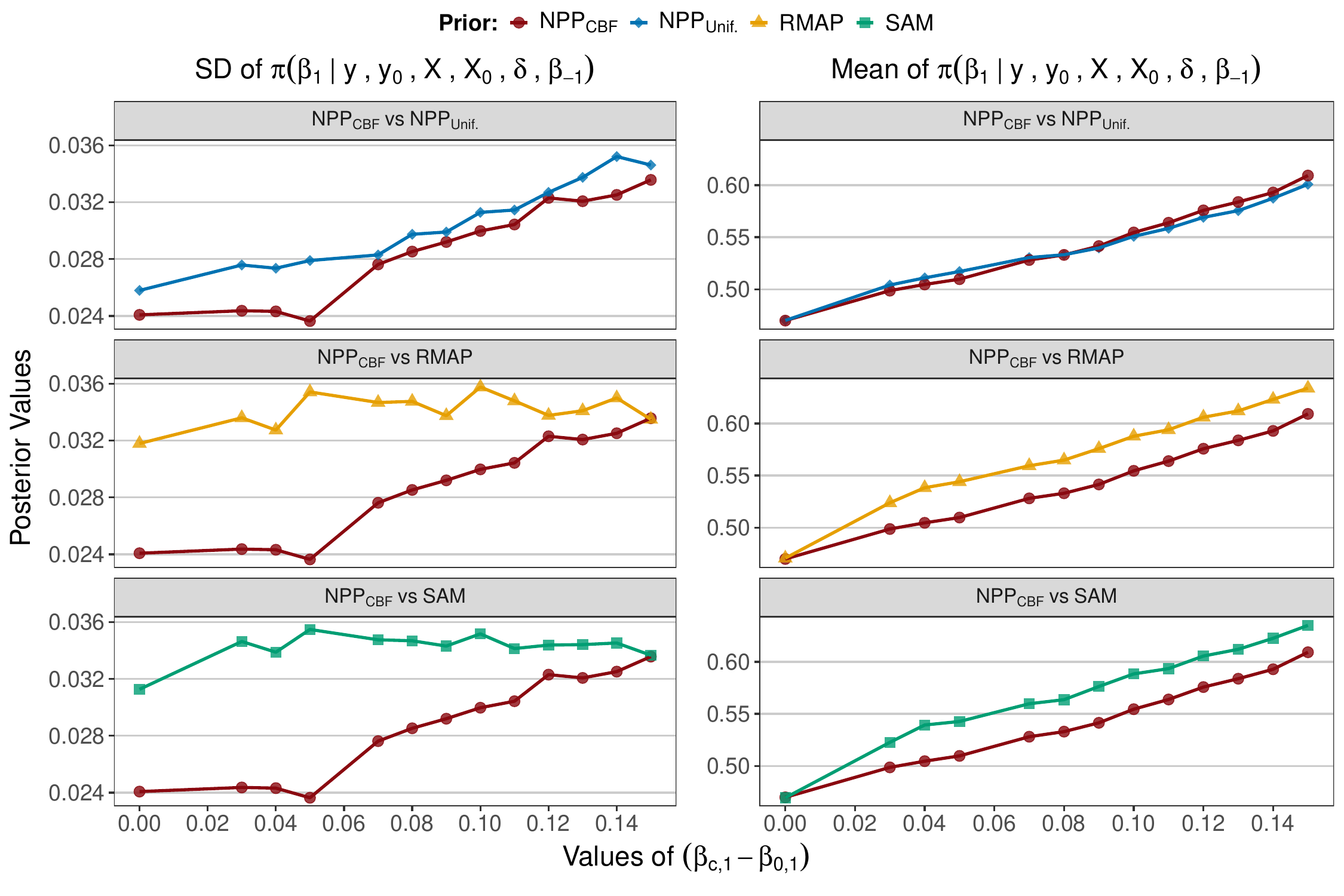}
        \caption{Poisson log-linear model. Standard deviation (SD) and mean of the marginal posterior for $\beta_1$, comparing four different priors: The normalized power prior (NPP) using the CBF derived Beta prior on $\delta$ (red dotted lines), the NPP using the standard uniform prior (blue diamond lines), the robust meta-analytic predicitve prior (RMAP) (yellow triangular lines), and the self-adapting mixture prior (SAM) (green squared lines). }
        \label{BF_plot_poisson_sd}
\end{figure}

Figure \ref{BF_plot_poisson_sd} shows the standard deviation (SD) and mean of the marginal posterior distribution for the regression coefficient $\beta_1$. The left panel of Figure \ref{BF_plot_poisson_sd}
illustrates that the marginal posterior distributions using the CBF selected prior are less diffuse than those derived from the standard uniform prior, the RMAP prior and the SAM prior, leading to more precise results for $\beta_1$. Since the SAM prior does not allow the incorporation of covariate information \citep{Yang_2023}, we compute the mixing weight for it using a Poisson model on the count outcome. Specifically, when there is minimal disagreement between current and historical data, the posterior standard deviation for $\beta_1$ is lower when using the CBF selected prior. As disagreement increases, the difference between the standard deviations tends to reduce. 
Furthermore, the right panel of Figure \ref{BF_plot_poisson_sd} highlights that the posterior mean in all the scenarios shows a similar increasing trend. This can be attributed to the progressive increase in the current regression coefficient $\beta_{c,1}$ which results in a greater discrepancy between current and historical data, leading to a higher discount of historical data. 

\begin{figure}[htb!]
    \centering
        \includegraphics[width=0.95\textwidth, height=8.5cm]{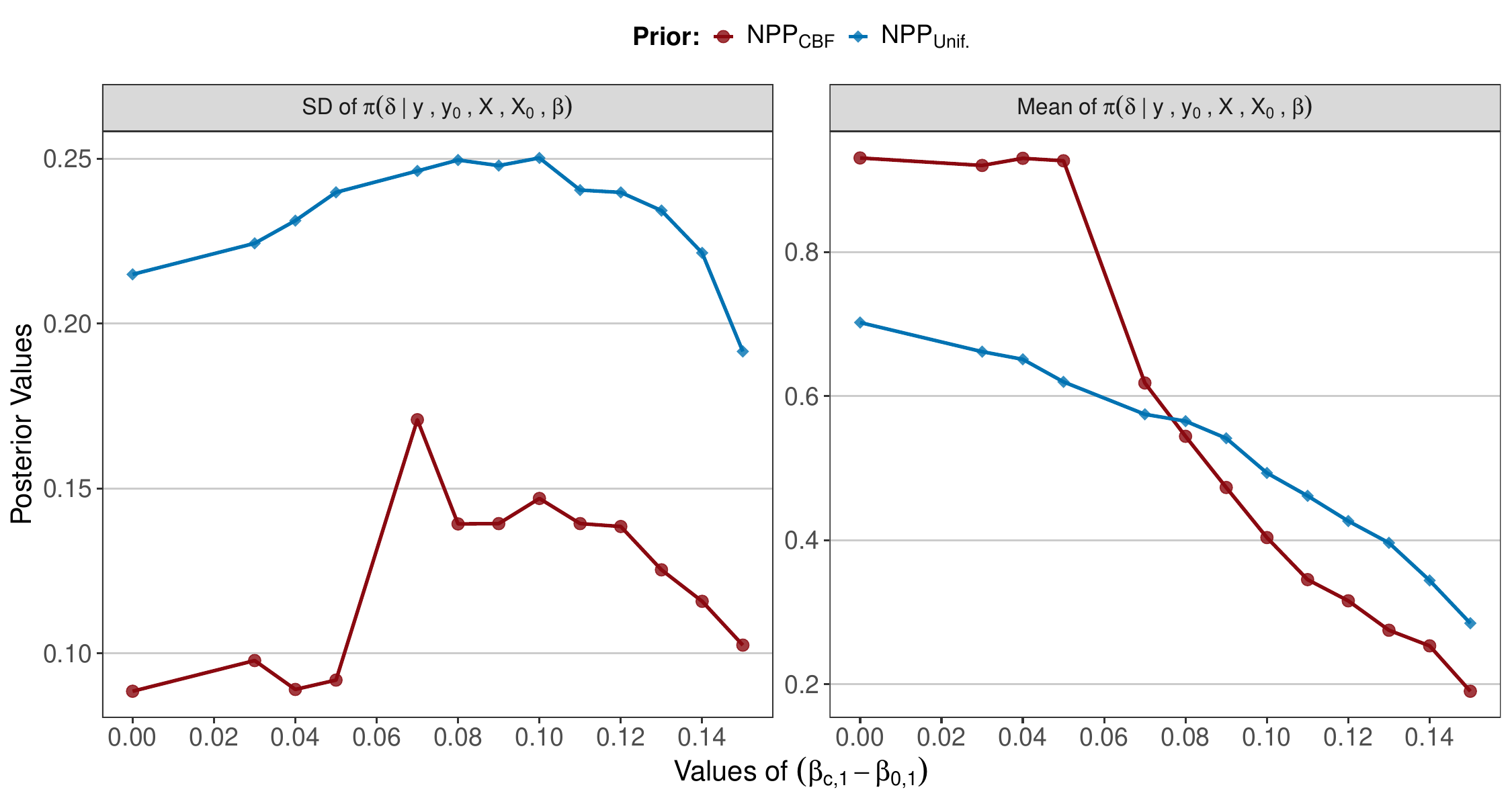}
        \caption{Poisson log-linear model. Standard deviation (SD) and mean of the marginal posterior for $\delta$ using the standard uniform prior (blue squared lines) and the CBF derived Beta prior (red dotted lines).}
        \label{GLM_poisson_delta_plot}
\end{figure}

Figure \ref{GLM_poisson_delta_plot} shows the posterior distribution of the weight parameter $\delta$. When there is a low level of disagreement between current and historical data, the CBF prior leads to posterior distributions that incorporate more historical information compared to the uniform prior for $\delta$. Conversely, as the disagreement increases, the CBF prior becomes more conservative, discounting the historical data to a greater extent. Furthermore, the CBF prior consistently leads to more precise estimates with lower variability in the posterior distribution.

\subsection{Binomial model}
A similar simulation study is conducted for the binomial model, which is frequently applied in medical contexts involving power priors. Let $N_0$ and $N$ denote the number of Bernoulli trials in the historical and current studies, respectively. Furthermore, let $y_0$ and $y$ represent the number of successes in these studies. Assuming a binomial likelihood with success probability $\theta$ for each study, and an initial Beta prior for both $\theta$ and the weight parameter $\delta$, the BF is
\begin{equation*}
    \mathrm{BF}_{0,i}(y) = \dfrac{\int_0^1 \mathrm{BBin}(y \mid N, \delta y_0 + p, \delta (N_0 - y_0) + q) \mathrm{Beta}\left(\delta \mid \eta_i, \nu_i\right) \mathrm{d} \delta}{\int_0^1 \mathrm{BBin}(y \mid N, \delta y_0 + p, \delta (N_0 - y_0) + q) \mathrm{Beta}\left(\delta \mid 1, 1\right) \mathrm{d} \delta}, \;\;\; i=1, \ldots, M,
\end{equation*}
where $\mathrm{BBin}(\cdot \mid N, \alpha, \beta)$ is the beta-binomial discrete distribution. For further details, see Appendix \ref{appendix_binomial}.

Figure \ref{BF_plot_binomial_opt} shows the evolution of the selected prior for $\delta$ when analyzing a historical dataset followed by a series of current datasets. A $\mathrm{Beta}(1,1)$ is used as the initial prior distribution for $\theta$, and the historical binomial likelihood presents a success probability of $0.2$. This simulation study demonstrates the ability of the proposed method to dynamically adapt the amount of historical information borrowed, based on the agreement between current and historical data. Specifically, when there is almost perfect agreement, the selected prior for $\delta$ is $\mathrm{Beta}(6,0.5)$, indicating a higher level of historical information borrowing. As the level of agreement decreases, the prior for $\delta$ progressively shifts to $\mathrm{Beta}$ distributions that reduce the incorporation of historical data, reaching a $\mathrm{Beta}(0.5,6)$ distribution in cases of high disagreement.
\begin{figure}[htb!]
    \centering
        \includegraphics[width=0.95\textwidth, height=8.5cm]{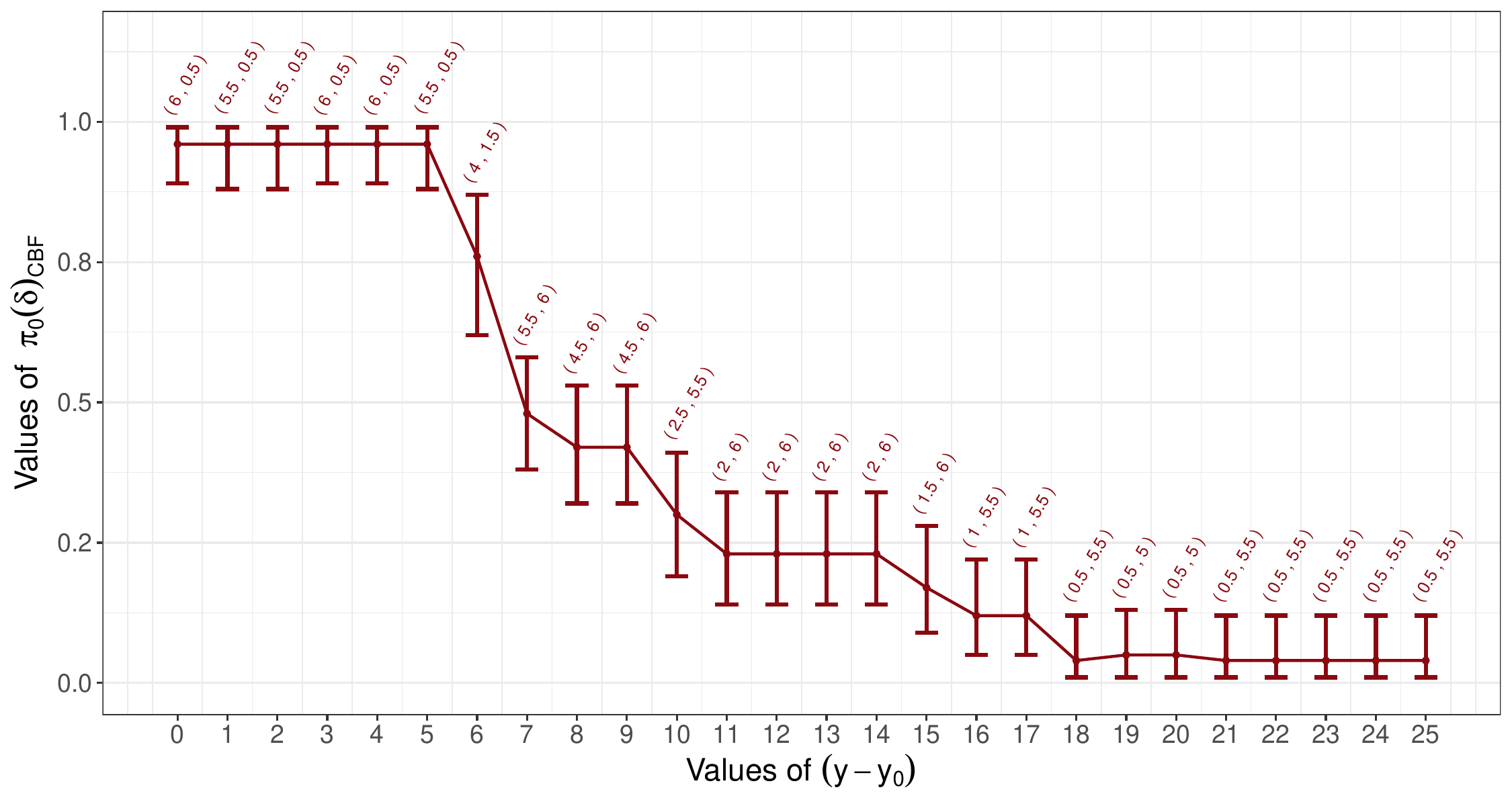}
        \caption{Binomial model. Shifts in the prior median (point) for the CBF derived prior for $\delta$, considering the 25th and 75th Percentiles (bar), in function of the difference between the current successes $y$ and the historical successes $y_0$. Above each interval, the corresponding selected Beta parameter are displayed.}
        \label{BF_plot_binomial_opt}
\end{figure}

\begin{figure}[htb!]
    \centering
        \includegraphics[width=0.95\textwidth, height=10.5cm]{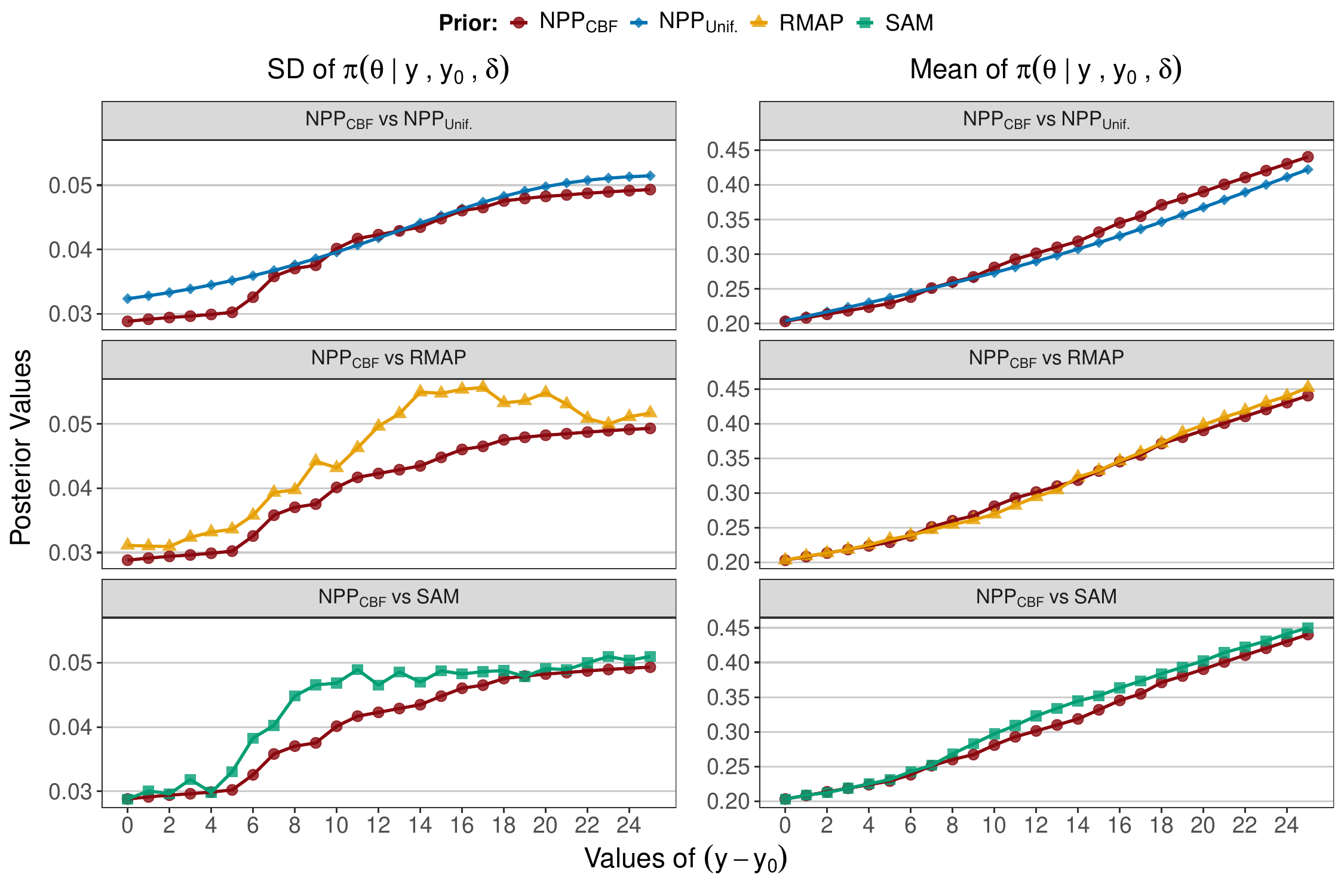}
        \caption{
        Binomial model. Standard deviation (SD) and mean of the marginal posterior for $\theta$, comparing four different priors: The normalized power prior (NPP) using the CBF derived Beta prior on $\delta$ (red dotted lines), the NPP using the standard uniform prior (blue diamond lines), the robust meta-analytic predicitve prior (RMAP) (yellow triangular lines), and the self-adapting mixture prior (SAM) (green squared lines).}
        \label{BF_plot_binomial_sd}
\end{figure}

Figure \ref{BF_plot_binomial_sd} presents the marginal posterior standard deviations and means for $\theta$ comparing the CBF selected initial prior for the weight parameter with the other dynamic information borrowing methods. Specifically, the left panel of Figure \ref{BF_plot_binomial_sd} highlights that the CBF prior for $\delta$ yields marginal posterior distributions for $\theta$ that are generally more concentrated. This is particularly evident when there is either a low or high level of disagreement between current and historical data, compared to the NPP with a uniform prior on $\delta$. Conversely, when compared to the RMAP prior and the SAM prior, the NPP using the CBF procedure produces less diffuse posterior distributions when there is a moderate level of disagreement between current and historical data.  Furthermore, the right panel of Figure \ref{BF_plot_binomial_sd} shows that the posterior mean of $\theta$ follows a consistently similar trend compared to results obtained using the competing information borrowing methods. Therefore, using the CBF prior leads to more accurate inferential conclusions in general.

\begin{figure}[htb!]
    \centering
        \includegraphics[width=0.99\textwidth, height=8cm]{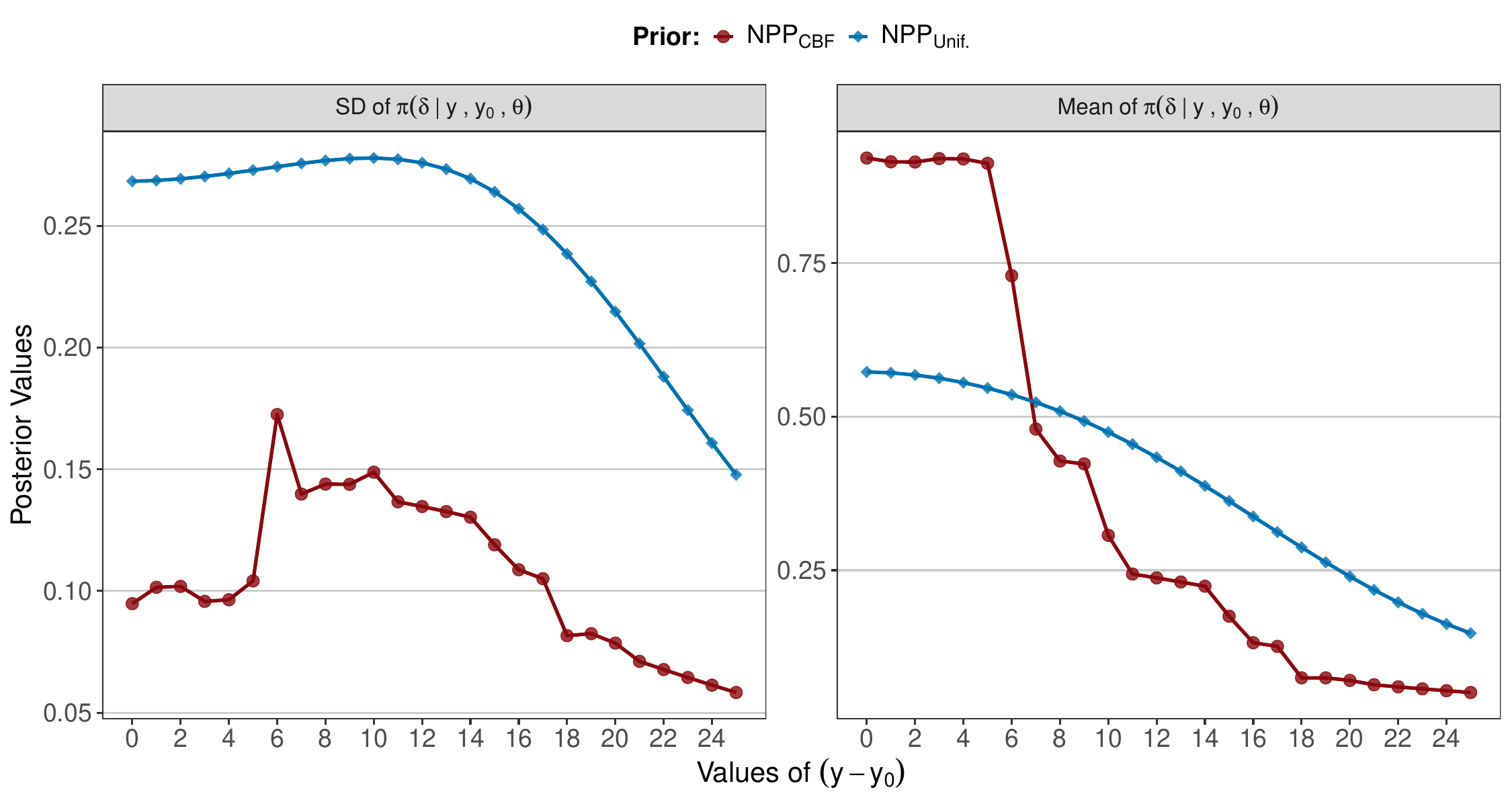}
        \caption{Binomial model. Standard deviation (SD) and mean of the marginal posterior for $\delta$ using the standard uniform prior (blue squared lines) and the CBF derived Beta prior (red dotted lines).}
        \label{binomial_delta_plot}
\end{figure}

Figure \ref{binomial_delta_plot} shows that when the discrepancy between the historical and current data is minimal, the CBF prior for $\delta$ produces posterior distributions that integrate more historical information. As the discrepancy increases, the posterior distributions become more conservative, progressively discounting the historical data. Furthermore, the CBF prior for $\delta$ generates less diffuse posterior distributions compared to those obtained from a uniform prior.

\subsection{Gaussian model}
We examine a Gaussian model commonly observed in replication studies \citep{pawel2023power}. 
A crucial question in this context is how effectively a replication (current) study has reproduced the results of an original (historical) study. Let $\mu$ denote the unknown true effect size, and let $\hat{\mu}_s$ represent the estimated effect size from study $s$, where $s \in \{o, r\}$ corresponds to the ``original'' and ``replication'' studies, respectively. Furthermore, the effect size estimates are assumed to be normally distributed.
\begin{align*}
    \hat{\mu}_s \mid \mu \sim \mathrm{N} \left(\mu, \sigma_s^2\right),
\end{align*}
where $\sigma^2_s$ represents the variance of the estimated effect size $\hat{\mu}_s$, assumed to be known. The BF is defined as
\begin{align*}
    \mathrm{BF}_{0,i}(y) &=\dfrac{\int_0^1 \mathrm{N} \left(\hat{\mu} \mid \hat{\mu}_o, \sigma^2+\sigma_o^2 / \delta\right) \mathrm{Beta}\left(\delta \mid \eta_i, \nu_i\right) \mathrm{d} \delta}{\int_0^1 \mathrm{N} \left(\hat{\mu} \mid \hat{\mu}_o, \sigma^2+\sigma_o^2 / \delta\right) \mathrm{Beta}\left(\delta \mid 1, 1\right) \mathrm{d} \delta}, \;\;\; i=1, \ldots, M.
\end{align*}
Further details are provided in Appendix \ref{appendix_normal}.

In Figure \ref{BF_plot_gaussian_mu_opt}, it is assumed that the effect size in the original study follows a normal distribution with mean $\mu_o = 0$ and variance $\sigma^2_o = 1$. We incrementally varied the true effect size of the replicated study in steps of 0.2, starting from a scenario of perfect agreement, where the replicated study's effect size is $\mu_r = 0$, and extending to scenarios of progressively greater disagreement, reaching a point where $\mu_r = 6$. Our method effectively selected a well-balanced prior to address the plausible level of agreement between the original and replicated studies. Similarly to the binomial case, a $\mathrm{Beta}(6,0.5)$ prior is chosen when the agreement is high. As disagreement increases, the amount of information borrowed is progressively reduced, selecting the prior that minimizes the incorporation of historical information -- a $\mathrm{Beta}(0.5,6)$ -- in cases of high disagreement. 
\begin{figure}[htb!]
    \centering
        \includegraphics[width=0.95\textwidth, height=8.5cm]{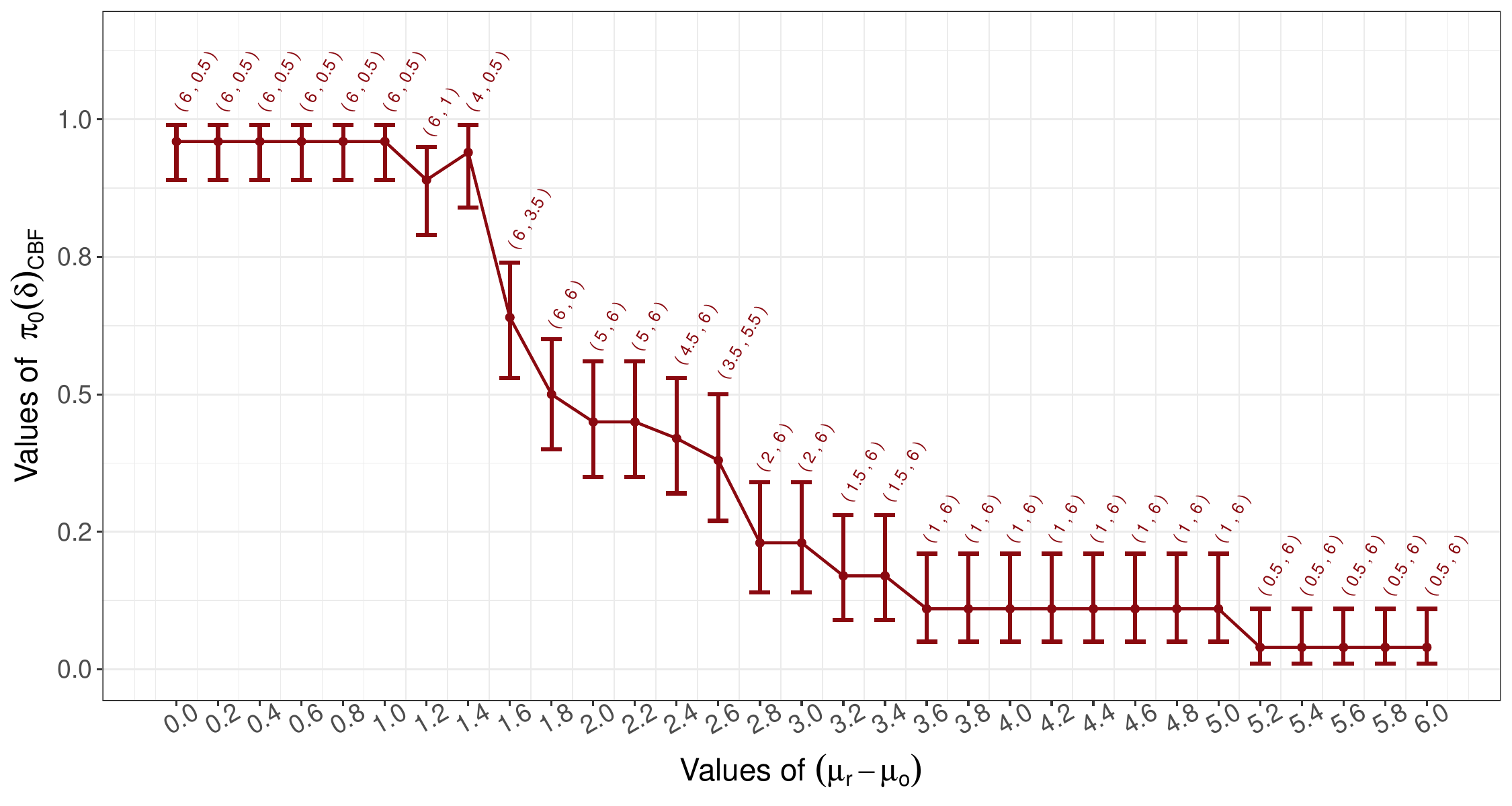}
        \caption{Gaussian model. Shifts in the prior median (point) for the CBF derived prior for $\delta$, considering the 25th and 75th Percentiles (bar), as a function of the difference between the replicated study's effect size $\mu_r$ and the original study's effect size $\mu_0$. Above each interval, the corresponding selected Beta parameter are displayed.}
        \label{BF_plot_gaussian_mu_opt}
\end{figure}

\begin{figure}[htb!]
    \centering
        \includegraphics[width=0.95\textwidth, height=10.5cm]{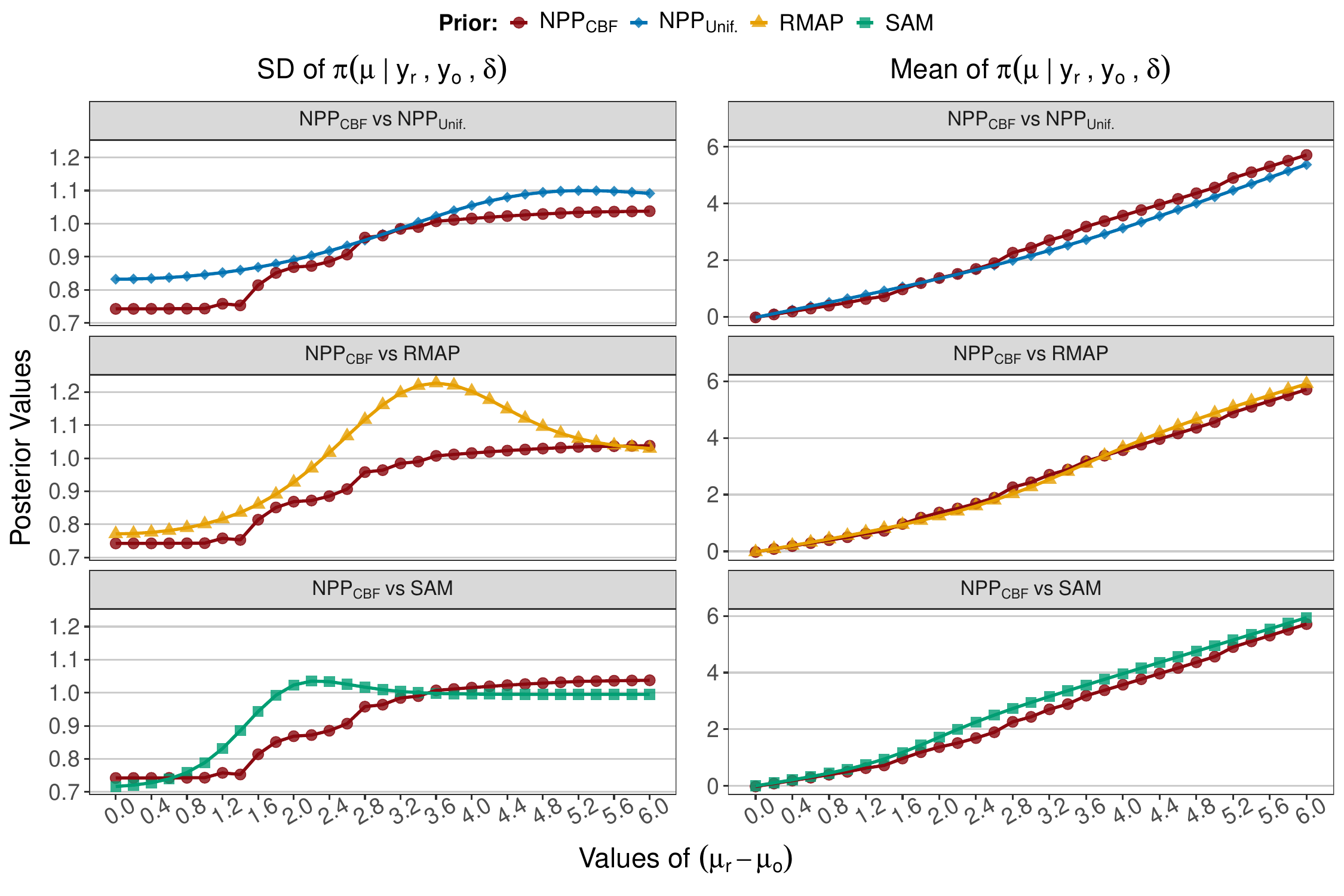}
        \caption{Gaussian model. Standard deviation (SD) and mean of the marginal posterior for $\mu$, comparing four different priors: The normalized power prior (NPP) using the CBF derived Beta prior on $\delta$ (red dotted lines), the NPP using the standard uniform prior (blue diamond lines), the robust meta-analytic predicitve prior (RMAP) (yellow triangular lines), and the self-adapting mixture prior (SAM) (green squared lines).}
        \label{combined_plot_gaussian}
\end{figure}

Figure \ref{combined_plot_gaussian} shows that the CBF procedure effectively selects a prior for the weight parameter, reducing the standard deviation of the marginal posterior for $\mu$ compared
to the NPP with a uniform prior on $\delta$ and the RMAP prior. The NPP with the CBF prior achieves a greater reduction in the standard deviation of the marginal posterior for $\mu$ than the SAM prior. However, when the level of disagreement is high or low, the behavior of the NPP with the CBF prior and the SAM prior becomes more comparable. As in previous simulation studies, the posterior mean for the effect size $\mu$ remains consistent in all the methods evaluated.

\begin{figure}[htb!]
    \centering
        \includegraphics[width=0.95\textwidth, height=8.5cm]{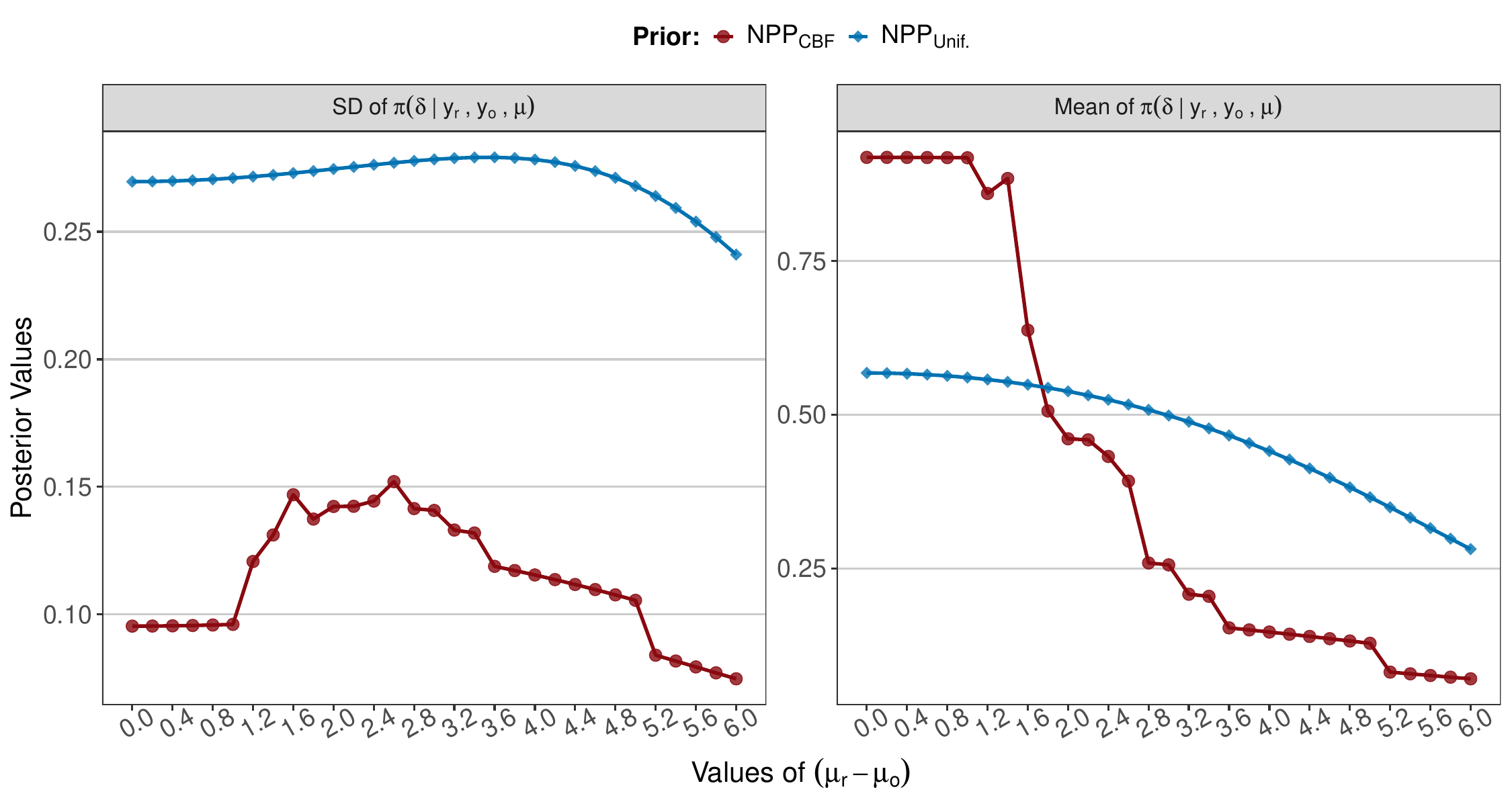}
        \caption{Gaussian model. Standard deviation (SD) and mean of the marginal posterior for $\delta$ using the standard uniform prior (blue squared lines) and the CBF derived Beta prior (red dotted lines).}
        \label{gaussian_delta_plot}
\end{figure}

Furthermore, Figure \ref{gaussian_delta_plot} shows results similar to those observed in previous simulation studies for the posterior distribution of $\delta$.

\section{Melanoma clinical trial} \label{Cancer Study}
\begin{figure}[htb!]
    \centering
        \includegraphics[width=0.95\textwidth, height=10.5cm]{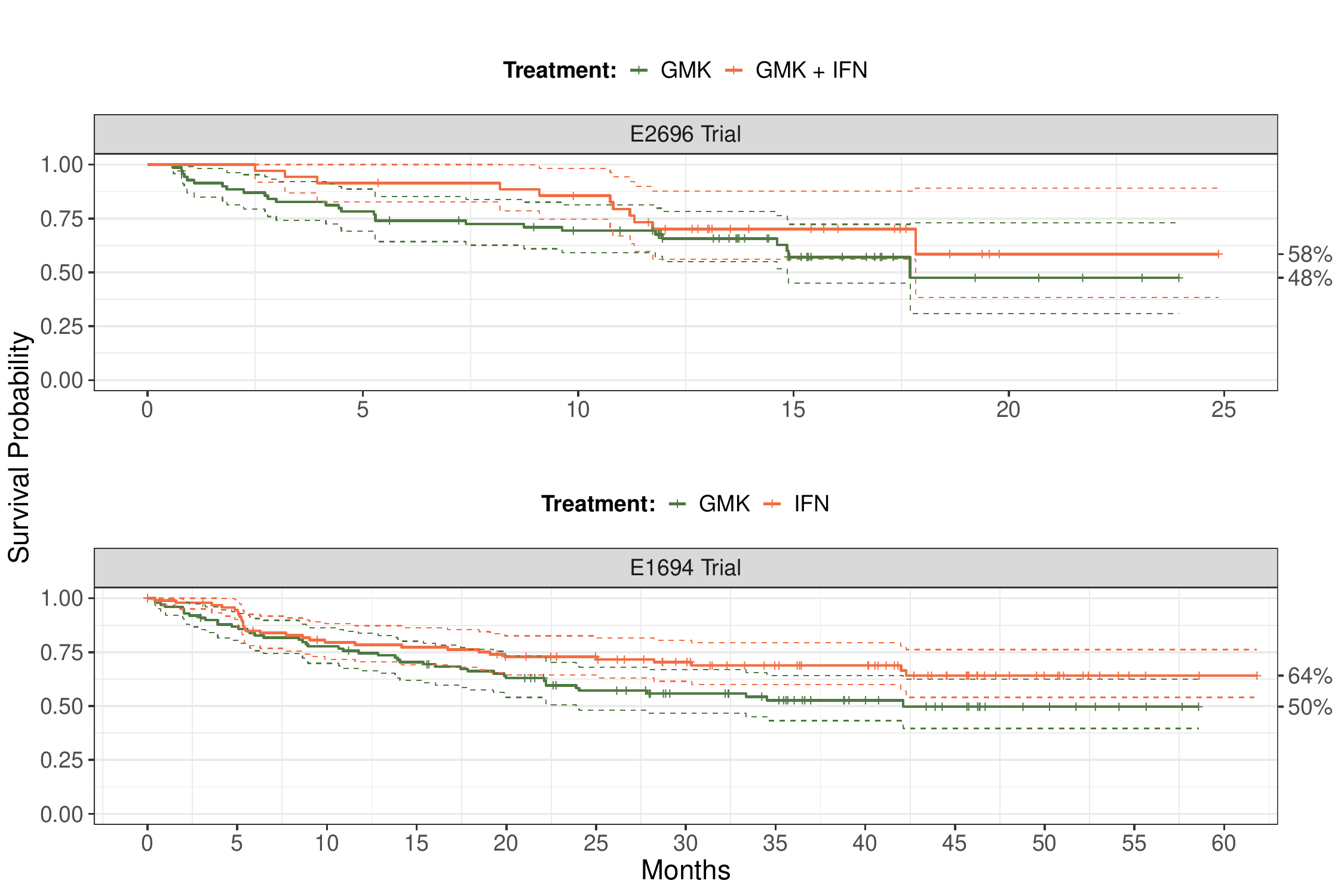}
        \caption{Melanoma clinical trial. Survival Curves for E2696 and E1694 Trials. On the right, the end time survival probability for each treatment. }
        \label{surv_fun_melanoma}
\end{figure}

The efficacy of the CBF procedure on real data is evaluated by analyzing two clinical trials. This analysis incorporates recent data from a new trial along with historical data from a previous study to assess earlier findings. Two melanoma trials conducted by the Eastern Cooperative Oncology Group (ECOG), specifically E2696 and E1694, are examined, involving 105 and 200 patients, respectively. For further details, see \citet{kirkwood2001E2696, kirkwood2001E1694}. These trials investigate the effects of interferon alfa-2b (IFN) treatment on patient survival rates. The E2696 trial evaluates the efficacy of combining the GM2-KLH/QS-21 (GMK) vaccine with high-dose IFN therapy compared to the GMK vaccine alone in patients with resected high-risk melanoma. Furthermore, the E1694 trial evaluates the effectiveness of the GMK vaccine versus high-dose IFN therapy in a comparable group of patients. In conclusion, the findings of the E1694 trial corroborate earlier results of E2696, demonstrating that intravenous and subcutaneous IFN can significantly reduce the relapse rate in patients with melanoma. Figure \ref{surv_fun_melanoma} presents the survival curves for both trials, highlighting the beneficial impact of interferon treatment on patient survival.

A Bayesian logistic regression model is applied to the data from the E1694 trial and additional historical information from the E2696 trial is integrated using a NPP as in (\ref{eqn: pp_normalized}). The analysis includes four additional covariates: age, sex, performance status, and treatment indicator. Parameter estimation is conducted using the probabilistic programming language Stan \citep{stan} to perform Markov Chain Monte Carlo (MCMC) sampling via the \texttt{hdbayes} R package \citep{hdbayes}. This involves four independent chains, each with 2000 iterations, discarding the first 1000 iterations as burn-in. To determine a well-balanced initial prior for the weight parameter using the CBF procedure, as described in Section \ref{subsec: The procedure}, the bridge sampling approximation of the BF is used. 

The initial priors for the coefficients of the four covariates are set using a weakly informative approach as outlined by \citet{Gelman2008}. Specifically, a normal distribution with mean 0 and standard deviation 10 is assigned to each coefficient. Furthermore, for the initial prior of the weight parameter, a $\mathrm{Beta}(5, 0.5)$ is chosen based on the CBF procedure. This process involves a comprehensive evaluation of competing initial priors, considering a range of different Beta parameters from 0.5 to 6 in increments of 0.5. Furthermore, a parallel processing strategy is employed to efficiently manage the computational effort required for this extensive parameter exploration, ensuring a streamlined and effective computational execution of the methodology.

Table~\ref{tab:post_table_melanoma_transposed} compares the posterior estimates for the regression parameters obtained using the NPP with different initial $\mathrm{Beta}$ priors for $\delta$, as well as other dynamic information borrowing methods. These estimates include posterior means, standard deviations, and $95\%$ HPDIs. The NPPs considered include the one based on the CBF criterion described in~(\ref{eqn:selection_criterion}), the standard uniform prior, Jeffreys' prior, and a mixture of two $\mathrm{Beta}$ priors for $\delta$ akin to a model averaging approach  \citep{rover_2019,best_2021,weru2024information}. In addition, we consider the RMAP prior, which places equal weight $\omega$ on both the non-informative and informative components derived from historical data; the SAM prior for a binary endpoint; and a fully Bayesian commensurate prior (CP)~\citep{hobbs2011hierarchical, Hobbs2012}, which accounts for uncertainty in the commensurability parameter and employs a spike-and-slab formulation~\citep{mitchell1988bayesian}, specified as a mixture of two half-normal priors~\citep{Hobbs2012}. Notably, the well-balanced prior identified using the CBF procedure -- a $\mathrm{Beta}(5, 0.5)$ -- results in consistently smaller posterior standard deviations for the treatment, sex, and performance status parameters compared to the other evaluated priors, indicating more precise inferential conclusions. Furthermore, the HPDIs for all the parameters of interest are narrower under the CBF selected prior, suggesting that, by incorporating substantial historical information, it enhances the precision of the posterior parameter estimation.
\begin{table}[htb!]
    \centering
    \caption{Melanoma clinical trial. Posterior mean, standard deviation (SD), and $95\%$ HPDI for the age, treatment, sex, and performance status comparing different initial prior for the weight parameter of the normalized power prior (NPP), the robust meta-analytic-predictive (RMAP) prior, self-adapting mixture (SAM) prior, and the commensurate prior (CP)}
    \begin{tabular}{l l r r l}
        \toprule  
        \textbf{Prior} & \textbf{Parameter} & \textbf{Mean} & \textbf{SD} & \textbf{95\% HPDI} \\
        \midrule
        \multicolumn{5}{l}{\textbf{NPP}} \\
        \midrule
        Beta(5, 0.5)$^{(1)}$  & Age    & 0.015 & 0.010 & (-0.004, 0.033) \\
                              & Treat. & -0.507 & 0.243 & (-0.999, -0.055) \\
                              & Sex    & -0.110 & 0.253 & (-0.600, 0.386) \\
                              & Perf.  & -0.462 & 0.326 & (-1.106, 0.156) \\
        \midrule
        Beta(1, 1)$^{(2)}$    & Age    & 0.015 & 0.010 & (-0.006, 0.033) \\
                              & Treat. & -0.525 & 0.267 & (-1.035, -0.013) \\
                              & Sex    & -0.138 & 0.276 & (-0.679, 0.374) \\
                              & Perf.  & -0.490 & 0.346 & (-1.135, 0.199) \\
        \midrule
        Beta(0.5, 0.5)$^{(3)}$& Age    & 0.015 & 0.010 & (-0.005, 0.034) \\
                              & Treat. & -0.525 & 0.257 & (-1.028, -0.016) \\
                              & Sex    & -0.124 & 0.264 & ( -0.634, 0.385) \\
                              & Perf.  & -0.479 & 0.350 & (-1.125, 0.252) \\
        \midrule
         $0.5\times$ Beta(0.5, 6) $+$  
        $0.5\times$ Beta(6, 0.5) & Age    & 0.015 & 0.010 & (-0.003, 0.036) \\
                                & Treat. & -0.517 & 0.250 & (-1.03, -0.042) \\
                                & Sex    & -0.110 & 0.254 & (-0.606, 0.396) \\
                                & Perf.  & -0.473 & 0.358 & (-1.181, 0.211) \\
        \midrule
        \multicolumn{5}{l}{\textbf{RMAP}} \\
        \midrule
        $\omega = 0.5$        & Age    & 0.015 & 0.010 & (-0.005, 0.034) \\
                              & Treat. & -0.554 & 0.284 & (-1.093, -0.065) \\
                              & Sex    & -0.183 & 0.268 & (-0.673, 0.368) \\
                              & Perf.  & -0.590 & 0.394 & (-1.296, 0.199) \\
        \midrule
        \multicolumn{5}{l}{\textbf{SAM}} \\
        \midrule
        $\omega = 0.96$       & Age    & 0.014 & 0.010 & (-0.006, 0.033) \\
                              & Treat. & -0.552 & 0.283 & (-1.086, -0.046) \\
                              & Sex    & -0.186 & 0.273 & (-0.745, 0.304) \\
                              & Perf.  & -0.589 & 0.408 & (-1.256, 0.279) \\
        \midrule
        \multicolumn{5}{l}{\textbf{CP}} \\
        \midrule
        Spike \& Slab         & Age    & 0.015 & 0.011 & (-0.004, 0.038) \\
                              & Treat. & -0.546 & 0.270 & (-1.063, -0.023) \\
                              & Sex    & -0.166 & 0.287 & (-0.724, 0.389) \\
                              & Perf.  & -0.502 & 0.395 & (-1.282, 0.251) \\
        \bottomrule
    \end{tabular} 
    \caption*{$^{(1)}$CBF, $^{(2)}$Uniform, $^{(3)}$Jeffreys}
    \label{tab:post_table_melanoma_transposed}
\end{table}

\begin{figure}[htb!]
    \centering
        \includegraphics[width=0.95\textwidth, height=10.5cm]{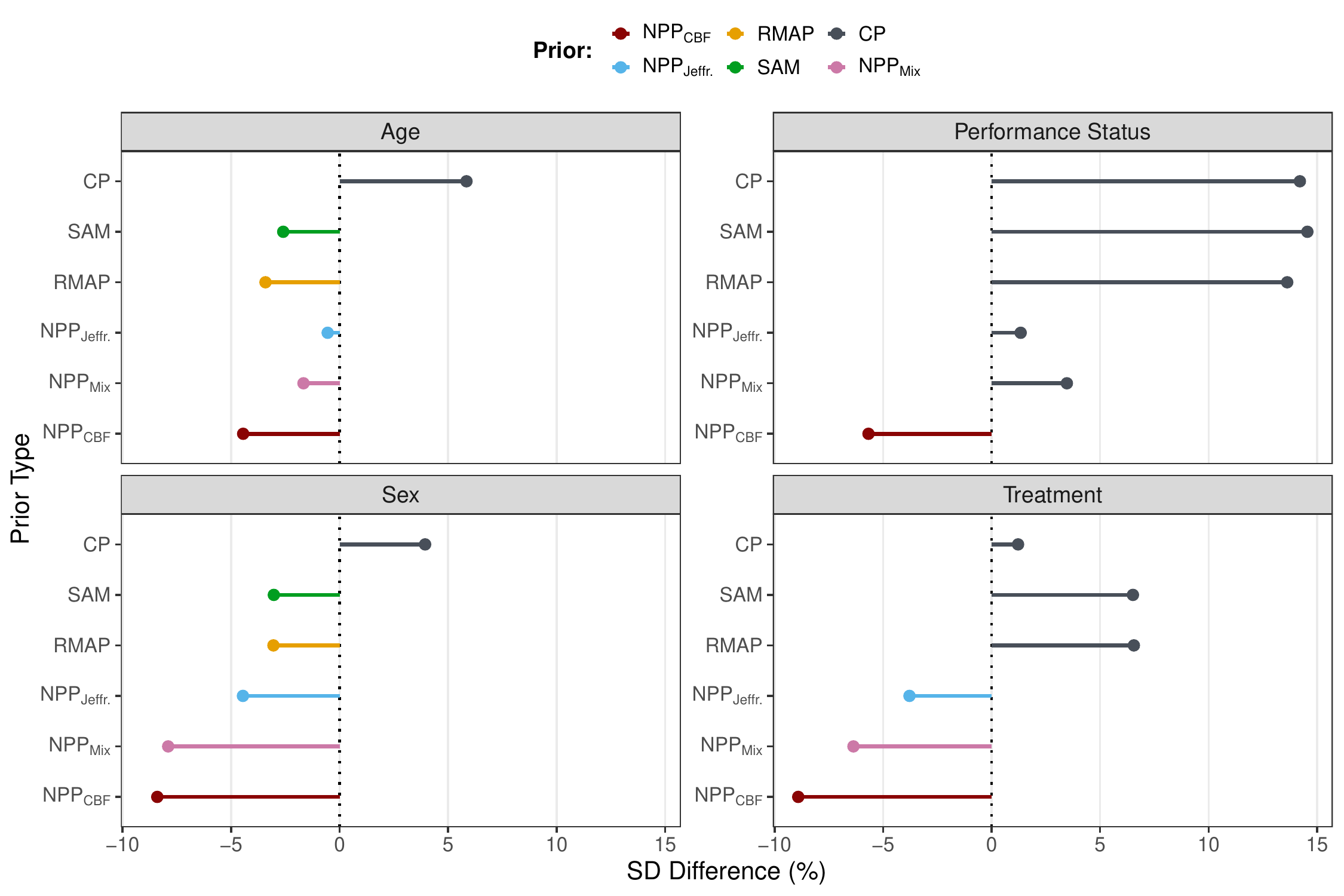}
        \caption{Melanoma clinical trial. Horizontal barplot for the age, performance status, sex, and treatment indicator representing the difference in percentage of the standard deviation of the evaluated priors in comparison with the normalized power prior with a uniform prior for $\delta$. }
        \label{lollipop_melanoma}
\end{figure}

Figure \ref{lollipop_melanoma} presents a horizontal bar plot that compares the standard deviations of the examined prior distributions with that of the NPP using a uniform prior on $\delta$. Among the evaluated priors, the NPP based on the CBF criterion demonstrates the most significant overall reduction in standard deviation for all parameters of interest.


\section{Discussion} \label{Discussion}
The power prior method presents a flexible way to construct an informative prior by combining a prior distribution with the weighted likelihood of previous data. This combined posterior then serves as the prior for new studies. However, determining the appropriate weight parameter presents a significant challenge, whether it is fixed or its prior distribution is being evaluated. Although several methods exist for setting a fixed weight parameter, fully Bayesian approaches for eliciting more informative priors are usually not addressed in the related literature.

\citet{Gravestock2017} highlighted that while the fully Bayesian approach is inherently flexible, it may not effectively address the disagreement between historical and current data. This issue frequently stems from the default use of a non-informative prior, which might not be sensitive enough to detect significant divergences. Consequently, we advocate for the use of a more informative prior specifically designed to detect potential conflict between historical and current datasets, improving the robustness of the resulting statistical inferences.

Our proposed CBF procedure is a novel response to this challenge. It seeks to select a more informative prior than the conventional non-informative one by using hypothetical replications derived from the posterior predictive distribution. The selected prior has minimal influence on the posterior central summary statistics while simultaneously achieving a smaller posterior variance for the parameters of interest. 
The efficacy of this approach is demonstrated through both simulation studies and the application to melanoma data, proving its robustness and effectiveness in distinguishing between different prior specifications. The ability of this method to select a well-balanced prior based on the agreement between historical and current data, as evidenced in the melanoma study, emphasizes its practical relevance in real-world applications. 
Furthermore, the flexibility of our proposal suggests that it can be easily extended to other types of endpoint, such as survival outcomes, adopting, for instance, the cure rate model proposed by \citet{Ibrahim2015}.

Our approach requires a higher computational cost to specify the weight parameter distribution, compared to simply assuming a standard uniform prior for $\delta$. This cost depends mainly on the formulation of the model, the number of hypothetical replications used for the log-BF distribution, and the grid search method. However, we believe that this additional effort is justified by more precise inferential conclusions about the parameters of interest. In practice, users can adapt the computational load to the formulation of their models by reducing the number of replications or using parallel computing, while still benefiting from the improved accuracy and robustness offered by the CBF procedure.

\color{black}

A crucial aspect of our CBF procedure is the choice of the HPDI to assess the location of the observed Log-BF within the distribution of replicated Log-BFs. This decision is crucial because it directly influences the interpretation of empirical evidence relative to the modeled hypotheses. A narrower HPDI is recommended when the goal is to limit the range of acceptable values, thereby enhancing the strength and reliability of empirical findings. Future research will focus on developing quantitative methods to determine the appropriate HPDI width.

The methodology presented in this paper offers several areas for potential improvement. Firstly, the selection criteria outlined in (\ref{eqn:selection_criterion}) could be refined to more effectively identify well-calibrated priors, particularly in cases of moderate agreement between historical and current data. Additionally, it is advisable to consider alternatives to the grid search method used in this study. Instead of exhaustively exploring all the parameters within the grid, methods that target a relevant subset of the parameter space should be explored. Furthermore, future work should focus on providing a more comprehensive analysis of the theoretical properties of the CBF method.

Finally, a thorough comparison with the optimal prior proposed by \citet{shen2023optimal} and methods that provide an estimate for $\delta$, possibly in terms of MSE or other measures, is a primary goal for future research.








\section*{Software and Data Availability}
All analyses were conducted in the R programming language version $4.4.1$ \citep{R_software_2023}. The code and data to reproduce this manuscript are openly available at \url{https://github.com/RoMaD-96/CBFpp}.

\bibliographystyle{apalike}

\bibliography{main}

\appendix

\section{Poisson log-linear model}
\label{appendix_poisson}
Let $y_0 = (y_{0,1}, \ldots, y_{0,N_0})$ and $y = (y_1, \ldots, y_N)$ be the count outcome of an historical and a current study, respectively. Let denote with $\boldsymbol{x}_{0,k}=(x_{0,k1}, \ldots, x_{0,kp})$, for $k = 1, \ldots, N_0$, and $\boldsymbol{x}_j=(x_{j1}, \ldots, x_{jp})$, for $j = 1, \ldots, N$, the corresponding covariate vector. The Poisson log-linear model is
\begin{align*}
y_j \mid \boldsymbol{\beta} \sim \mathrm{Poisson}(\lambda_j), \quad \text{where } \lambda_j = \exp(\mathbf{x}_j^\top \boldsymbol{\beta}),   
\end{align*}
with $\boldsymbol{\beta}$ being the $p$-dimensional vector of the regression coefficients.
Let $\pi_0(\boldsymbol{\beta})$ be an initial multivariate normal prior on $\boldsymbol{\beta}$
\begin{align*}
\boldsymbol{\beta} \sim \mathrm{N}_p(\boldsymbol{m}, V),   
\end{align*}
where $\boldsymbol{m}$ is a $p$-dimensional mean vector and $V$ is a $p \times p$ covariance matrix.
The NPP for $(\boldsymbol{\beta}, \delta)$ is then
\begin{align*}
\pi(\boldsymbol{\beta}, \delta \mid y_0, X_0) = \frac{[L(\boldsymbol{\beta} \mid y_0, X_0)]^\delta \pi_0(\boldsymbol{\beta})}{\int [L(\boldsymbol{\beta} \mid y_0, X_0)]^\delta \pi_0(\boldsymbol{\beta}) \, \mathrm{d}\boldsymbol{\beta}} \times \mathrm{Beta}(\delta \mid \eta,\nu),    
\end{align*}
where
$[L(\boldsymbol{\beta} \mid y_0, X_0)]^\delta = \exp\left(\sum_{k=1}^{N_0} \left[\delta(y_{0,k}\mathbf{x}_{0,k}^\top \boldsymbol{\beta} - \exp(\mathbf{x}_{0,k}^\top \boldsymbol{\beta}))\right]\right)
\prod_{k=1}^{N_0} (y_{0,k}!)^{-\delta}.$
Given current data, the posterior distribution is
\begin{align*}
\pi(\boldsymbol{\beta}, \delta \mid y, y_0, X, X_0) &= \frac{L(\boldsymbol{\beta} \mid y, X)\pi(\boldsymbol{\beta}, \delta \mid y_0, X_0)}{\int_0^1 \int L(\boldsymbol{\beta} \mid y, X)\pi(\boldsymbol{\beta}, \delta \mid y_0, X_0)\, \mathrm{d}\boldsymbol{\beta} \, \mathrm{d}\delta}.
\\ &=  \frac{L(\boldsymbol{\beta} \mid y, X)[L(\boldsymbol{\beta} \mid y_0, X_0)]^\delta \pi_0(\boldsymbol{\beta})\mathrm{Beta}(\delta \mid \eta,\nu)}{\int_0^1 \int L(\boldsymbol{\beta} \mid y, X)[L(\boldsymbol{\beta} \mid y_0, X_0)]^\delta \pi_0(\boldsymbol{\beta}) \mathrm{Beta}(\delta \mid \eta,\nu)\, \mathrm{d}\boldsymbol{\beta} \, \mathrm{d}\delta},    
\end{align*}
where $L(\boldsymbol{\beta} \mid y, X) = \exp\left(\sum_{j=1}^{N} \left[(y_{j}\mathbf{x}_{j}^\top \boldsymbol{\beta} - \exp(\mathbf{x}_{j}^\top \boldsymbol{\beta}))\right]\right)
\prod_{j=1}^{N} y_{j}!.$
Consequently, the BF is
\begin{align*}
    \mathrm{BF}_{0,i}(y) = \frac{\int_0^1 \int L(\boldsymbol{\beta} \mid y, X)[L(\boldsymbol{\beta} \mid y_0, X_0)]^\delta \pi_0(\boldsymbol{\beta})\mathrm{Beta}(\delta \mid \eta_i,\nu_i) \, \mathrm{d}\boldsymbol{\beta} \, \mathrm{d}\delta}{\int_0^1 \int L(\boldsymbol{\beta} \mid y, X)[L(\boldsymbol{\beta} \mid y_0, X_0)]^\delta \pi_0(\boldsymbol{\beta})\mathrm{Beta}(\delta \mid 1,1) \, \mathrm{d}\boldsymbol{\beta} \, \mathrm{d}\delta}.
\end{align*}

\section{Binomial with unknown success probability $\theta$} \label{appendix_binomial}
Let $N_0$ and $N$ denote the number of Bernoulli trials in the historical and current studies, respectively. The terms $y_0$ and $y$ represent the successes in these studies. Assuming a binomial likelihood with a success probability $\theta$ for each study and an initial Beta prior for both $\theta$ and the weight parameter $\delta$, the NPP is 
\begin{equation*}
\begin{aligned}
   \pi(\theta, \delta \mid y_0) &= \dfrac{\left[\mathrm{Bin}(y_0 \mid \theta, N_0)\right]^{\delta} \mathrm{Beta}(\theta \mid p,q) }{\int_0^1 \left[\mathrm{Bin}(y_0 \mid \theta, N_0)\right]^{\delta} \mathrm{Beta}(\theta \mid p,q) \mathrm{d}\theta} \times \mathrm{Beta}(\delta \mid \eta, \nu)
    \\
    &= \mathrm{Beta}\left(\theta \mid \delta y_0 + p, \delta \left(N_0-y_0\right) + q \right)\mathrm{Beta}(\delta \mid \eta, \nu).
\end{aligned}
\label{eqn: npp_binomial}
\end{equation*}
In light of the current data, the posterior distribution is 
\begin{align*}
    \pi\left(\theta, \delta \mid y, y_0\right)&=\frac{L(\theta \mid y) \pi\left(\theta, \delta \mid y_0\right)}{\int_0^1 \int_{0}^{1} L\left(\theta \mid y\right) \pi\left(\theta, \delta \mid y_0\right) \mathrm{d} \theta \mathrm{d} \delta} 
    \\
    &=\dfrac{\mathrm{Bin}\left(y \mid \theta, N \right) \mathrm{Beta}\left(\theta \mid \delta y_0 + p, \delta \left(N_0-y_0\right) + q \right)\mathrm{Beta}(\delta \mid \eta, \nu)}{\int_0^1 \mathrm{BBin}(y \mid N, \delta y_0 + p, \delta (N_0 - y_0) + q) \mathrm{Beta}(\delta \mid \eta, \nu) \mathrm{d} \delta},
\end{align*}
where $\mathrm{BBin}(\cdot \mid N, \alpha, \beta)$ is the beta-binomial discrete distribution. Therefore, the BF is 
\begin{align*}
    \mathrm{BF}_{0,i}(y) &= \dfrac{\int_0^1 \mathrm{BBin}(y \mid N, \delta y_0 + p, \delta (N_0 - y_0) + q) \mathrm{Beta}\left(\delta \mid \eta_i, \nu_i\right) \mathrm{d} \delta}{\int_0^1 \mathrm{BBin}(y \mid N, \delta y_0 + p, \delta (N_0 - y_0) + q) \mathrm{Beta}\left(\delta \mid 1, 1\right) \mathrm{d} \delta}, \;\;\; i=1, \ldots, M.
\end{align*}
\color{black}

\section{Gaussian with unknown mean $\mu$} \label{appendix_normal}
Let $\mu$ be the unknown true effect size, with $\hat{\mu}_s$ representing the estimated effect size from study $s$, where $s \in \{o, r\}$ denotes ``original'' and ``replication'' studies, respectively. Furthermore, we assume that the effect size estimates are normally distributed.
\begin{align*}
    \hat{\mu}_s \mid \mu \sim \mathrm{N} \left(\mu, \sigma_s^2\right),
\end{align*}
where $\sigma^2_s$ represents the variance of the estimated effect size $\hat{\mu}_s$, assumed to be known. Let consider an initial improper prior for the effect size parameter $\pi_0(\mu)\propto 1$ and a Beta prior for the weight parameter $\delta$ then the NPP is
\begin{align*}
    \pi\left(\mu, \delta \mid y_o\right)&= \mathrm{N}\left(\mu \mid \hat{\mu}_o, \dfrac{\sigma_o^2}{\delta}  \right) \mathrm{Beta}(\delta \mid \eta, \nu).
\end{align*}
Updating the previous prior with the likelihood of the replicated data yields the following posterior distribution
\begin{align*}
    \pi\left(\mu, \delta \mid y, y_o\right)&=\frac{L(\mu \mid y) \pi\left(\mu, \delta \mid y_o\right)}{\int_0^1 \int_{-\infty}^{\infty} L\left(\mu \mid y\right) \pi\left(\mu, \delta \mid y_o\right) \mathrm{d} \mu \mathrm{d} \delta} 
    \\
    &=\dfrac{\mathrm{N}\left(\hat{\mu} \mid \mu, \sigma^2\right) \mathrm{N}\left(\mu \mid \hat{\mu}_o, \sigma_o^2 / \delta\right) \mathrm{Beta}(\delta \mid \eta, \nu)}{\int_0^1 \mathrm{N} \left(\hat{\mu} \mid \hat{\mu}_o, \sigma^2+\sigma_o^2 / \delta\right) \mathrm{Beta}\left(\delta \mid \eta, \nu\right) \mathrm{d} \delta}.
\end{align*}
Furthermore, the BF is
\begin{align*}
    \mathrm{BF}_{0,i}(y) &=\dfrac{\int_0^1 \mathrm{N} \left(\hat{\mu} \mid \hat{\mu}_o, \sigma^2+\sigma_o^2 / \delta\right) \mathrm{Beta}\left(\delta \mid \eta_i, \nu_i\right) \mathrm{d} \delta}{\int_0^1 \mathrm{N} \left(\hat{\mu} \mid \hat{\mu}_o, \sigma^2+\sigma_o^2 / \delta\right) \mathrm{Beta}\left(\delta \mid 1, 1\right) \mathrm{d} \delta}, \;\;\; i=1, \ldots, M.
\end{align*}

\end{document}